\UseRawInputEncoding
\documentclass[journal]{IEEEtran}

\usepackage{cite}

\ifCLASSINFOpdf
    \usepackage[pdftex]{graphicx}
    \graphicspath{{../figures}}
\else
\fi
\usepackage{amsmath}
\usepackage{multirow}

\usepackage{hhline}
\usepackage{textcomp}
\usepackage{multicol}
\usepackage{csquotes}
\begin{document}
%
\title{Analysis and Modeling of an 11.8 GHz Fin Resonant Body Transistor in a 14nm FinFET CMOS Process}
%
%
%

\author{Udit Rawat,~\IEEEmembership{Student~Member,~IEEE,}
        Bichoy Bahr,~\IEEEmembership{Member,~IEEE,}
        Dana~Weinstein,~\IEEEmembership{ Senior~Member,~IEEE}
\thanks{This work was supported in part by the DARPA MIDAS Program. }
\thanks{Udit Rawat is with the Department of Electrical Engineering, Purdue University, West Lafayette, IN 47906 USA (e-mail: rawatu@purdue.edu).}
\thanks{Bichoy Bahr is with Kilby Labs - Texas Instruments, Dallas, TX, USA.}
\thanks{Dana Weinstein is with the Department of Electrical Engineering, Purdue University, West Lafayette, IN 47906 USA (e-mail: danaw@purdue.edu).}
}

\maketitle

\begin{abstract}
In this work, a compact model is presented for a 14 nm CMOS-based FinFET Resonant Body Transistor (fRBT) operating at a frequency of 11.8 GHz and targeting RF frequency generation/filtering for next generation radio communication, clocking, and sensing applications. Analysis of the phononic dispersion characteristics of the device, which informs the model development, shows the presence of polarization exchange due to the periodic nature of the back-end-of-line (BEOL) metal PnC. An eigenfrequency-based extraction process, applicable to resonators based on electrostatic force transduction, has been used to model the resonance cavity. Augmented forms of the BSIM-CMG (Common Multi-Gate) model for FinFETs are used to model the drive and sense transistors in the fRBT. This model framework allows easy integration with the foundry-supplied process design kits (PDKs) and circuit simulators while being flexible towards change in transduction mechanisms and device architecture. Ultimately, the behaviour is validated against RF measured data for the fabricated fRBT device under different operating conditions, leading to the demonstration of the first complete model for this class of resonant device integrated seamlessly in the CMOS stack.
\end{abstract}


%
\IEEEpeerreviewmaketitle

\section{Introduction}
\label{sec:introduction}

\IEEEPARstart{I}{ncreasing} proliferation of cm- and mm-Wave 5G mobile communication technology to address the demand for high data rates, better reliability and low-latency necessitates innovation in the field of front-end electroacoustic devices for filtering and carrier generation. AlN and LiNbO$_{3}$ bulk acoustic wave (BAW) resonators and filters utilizing higher modes of operation have recently shown promise \cite{Rinaldi_AlN_COR}\cite{Gong_LiNbO3_10to60GHz} with regards to applicability in the currently allocated n257-n260 (24.5 to 40 GHz) 5G mm-Wave bands. However, these resonators have specialized fabrication and packaging requirements making their integration with CMOS prohibitive even with MEMS-last \cite{AmitLal_AlScN_MEMSlast}, Front-end-of-line (FEOL) \cite{Barniol} and  Back-end-of-line (BEOL) post-processed MEMS \cite{SSLi_CMOSMEMS}\cite{Fedder}\cite{Weileun} approaches. The typical solution of packaging the MEMS die separately from CMOS exacerbates the problem because of interconnect bandwidth limitations across multiple chips at cm- and mm-Wave frequencies.  Moreover, at these frequencies, carrier generation and distribution to all the channels in a multi-element phased array IC via the conventional off-chip crystal and PLL combination, as shown in \cite{BSadhu_PTPLL}, results in significant routing power dissipation. Availability of high-Q, integrated CMOS-MEMS resonator-based oscillators with good phase noise levels at the targeted frequencies would result in a reduction in the carrier power since the central PLL would no longer be necessary. Considering the aforementioned challenges and potential opportunities at the aforementioned high frequencies, monolithic integration of RF/mmWave MEMS resonators in a conventional CMOS process becomes an attractive proposition.  

Fully-integrated, solid-state, CMOS-MEMS RF/mmWave resonators have previously been demonstrated using different technology nodes \cite{bahr_theory_2015}\cite{bahr_32ghz_2018}. These resonators make use of acoustic waveguiding confinement based on Back-end-of-Line (BEOL) metal phononic crystals (PnCs) and adiabatic terminations for mode localization to form a resonance cavity concentrated at the transistor layers of the CMOS stack.  The targeted mode is excited differentially using metal-oxide-semiconductor (MOS) capacitors, or MOSCAPs, and sensed using a pair of transistors incorporated into the resonant cavity, biased in saturation. The drain current in these sense FETs is modulated by the stress generated in the channel due to vibration, resulting in a differential readout. Since these resonators are fully integrated within a given CMOS technology, high performance oscillators can be designed (e.g. \cite{Abhishek_Osc}) with significant savings in terms of carrier power generation and distribution as well as area and cost. To design and correctly estimate the performance of such oscillators, precise, physics-accurate compact models are required for the constituent CMOS-MEMS resonant devices. These models are expected to capture the coupled physics of the various transduction mechanisms as well as the mechanical behaviour of the device without having to resort to computationally intensive Finite Element Method (FEM) simulations.

An initial circuit model for a transistor-sensed CMOS-MEMS resonator, as depicted in \cite{Radhika_UnreleasedRes32nm} has previously been devised to capture the small-signal behaviour of a Resonant Body Transistor based on a simplified analytical model of the drive capacitor and sense transistor. In that model, the passive section of the device consists of a drive MOSCAP in accumulation and a cavity whose resonant behaviour is captured using a series resistance (R), inductance (L) and capacitance (C) equivalent circuit. The transistor sensing is implemented using an electromechanical transconductance $g_{m,em}$ which represents the transduction from the cavity mechanical resonance to drain current modulation. This model explains the fundamental behaviour of the resonator, but is not sufficiently detailed and suffers from two significant drawbacks. First, owing to the low motional capacitance of these devices with respect to the static drive capacitance $c_{0}$, it is very challenging to extract the equivalent R, L, and C values for the passive section experimentally at high frequencies. Second, oscillators typically operate in the regime of large signal swings to achieve low phase noise, so a small signal equivalent circuit does not capture the effect of various nonlinearities inherent to the device. 

\begin{figure}[t]
\centering
\includegraphics[scale=0.26]{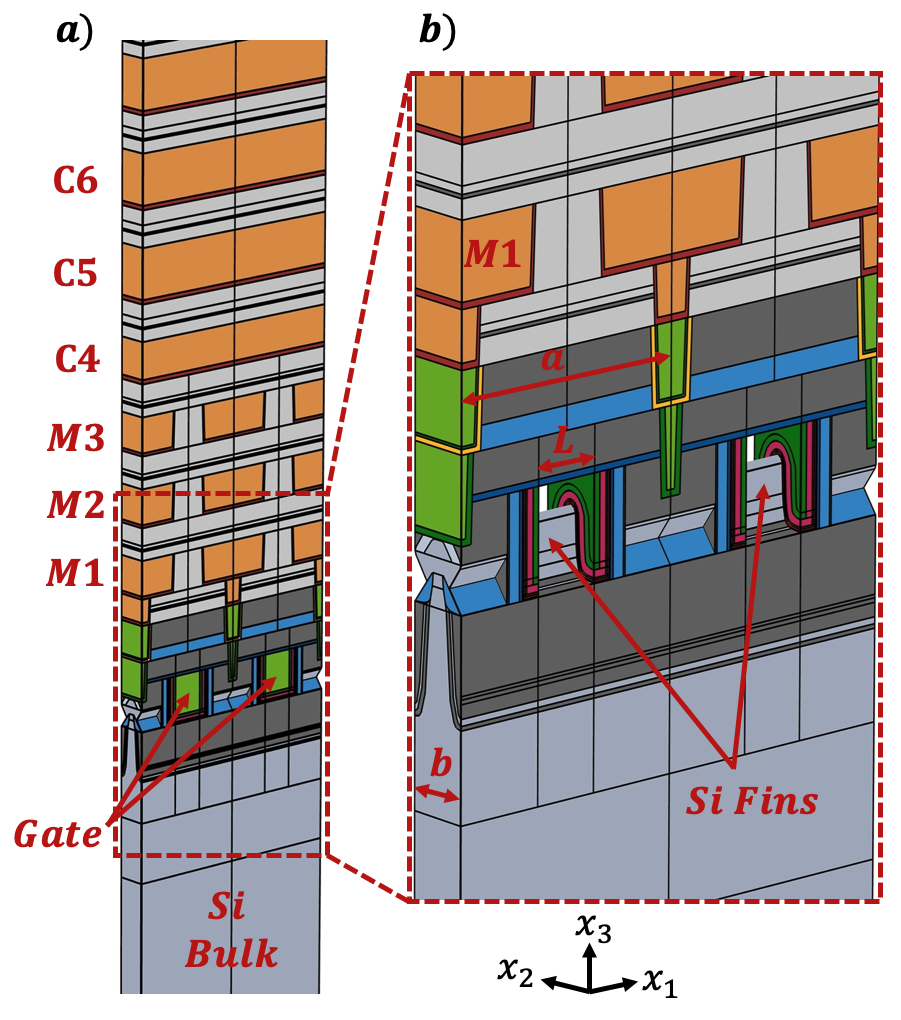}
\vspace{-2mm}
\caption{(a) Complete 3D FEM Simulation model depicting two adjoining fRBT unit cells. Mx (x=1-3) and Cy (y=4-6) represent the first 6 metal levels that form a part of the BEOL PnC. (b) Inset showing the front-end-of-line (FEOL) channel region with the gate stack hidden. The quantities $a$, $b$ and $L$ represent lattice constants in the $x_{1}$ and $x_{2}$ directions and the gate length respectively.}
\label{fig1}
\vspace{-4mm}
\end{figure}

 Compact models were subsequently developed for a 1-D Unreleased \cite{Bichoy_RBTMVSnanoHUB} and released \cite{Bichoy_RBTBSIMnanoHUB} Resonant Body Transistor (RBT)  which use modified MIT Virtual Source and BSIM planar FET models for the sense transistors only. While these model are large-signal in nature and improve upon the drawbacks of the small-signal model in \cite{Radhika_UnreleasedRes32nm}, they do not translate well to CMOS-integrated RBTs built using FinFETs as in \cite{bahr_32ghz_2018} and the resonator considered in this paper. The primary reason for this being the 3D nature of the silicon fin together with the complex mode shape requiring additional analysis for the modeling of various effects of the stress developed in the sense transistor channel. The model in \cite{Bichoy_RBTMVSnanoHUB} considers drain current modulation in the sense transistor only due to mobility modulation. However, in the actual device, stress and strain in the transistor channel cause changes to multiple other device parameters which need to be considered. These models also do not consider electrostatic drive using MOS capacitors. For the complete device model to be compatible with Electronic Design Automation (EDA) tools and the foundry-supplied process design kits (PDKs) used for simulation and design, the industry standard BSIM-CMG model \cite{BSIM-CMG} is required to be augmented to capture these effects accurately. 

\begin{figure}
    \centering
    \includegraphics[scale=0.167]{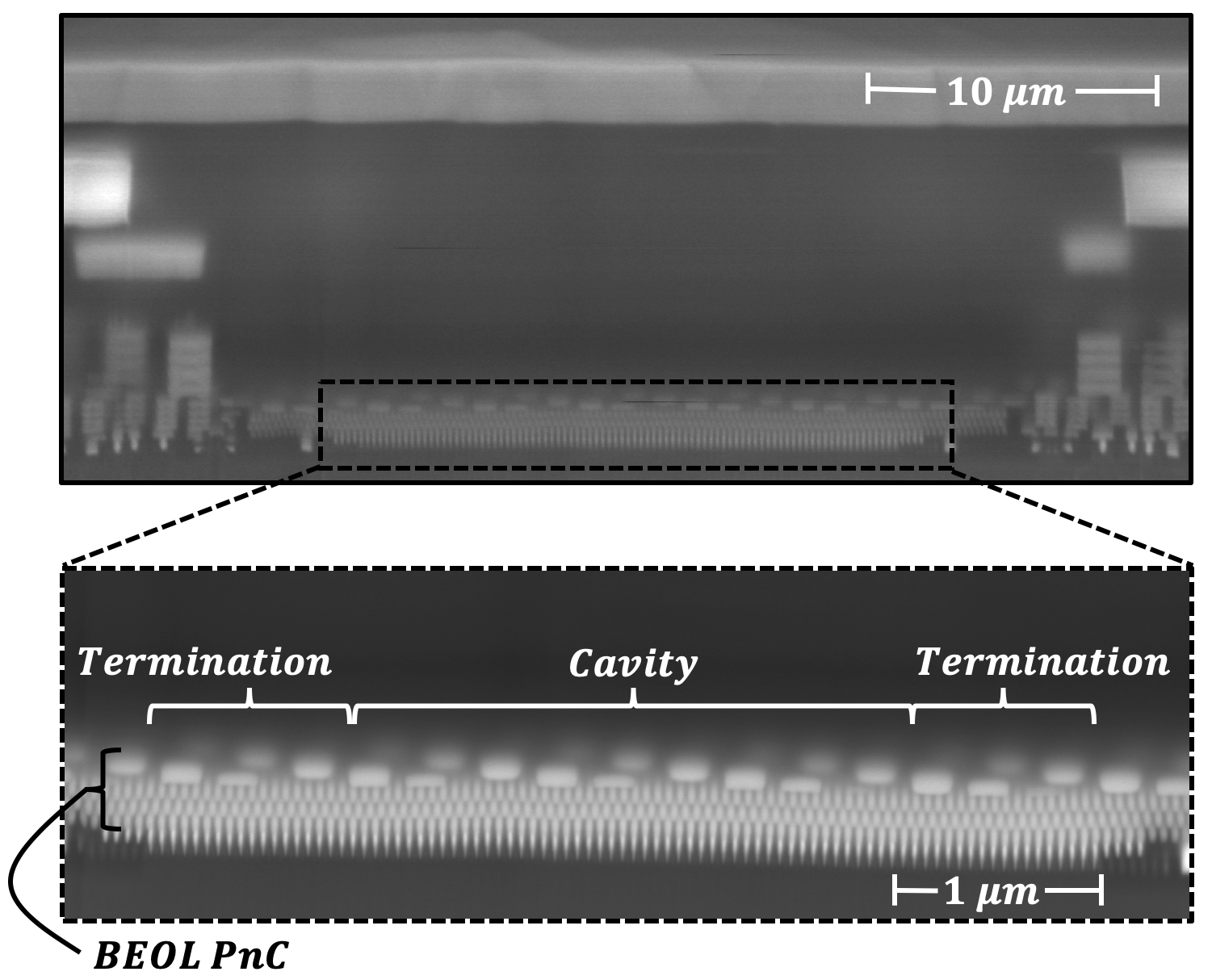}
    \caption{Cross-sectional SEM of the 11.8 GHz fRBT resonator depicting the resonant cavity bound laterally by the termination regions. The BEOL metal phononic crystal above the gate region can also be seen.}
    \label{fig11}
    \vspace{-3mm}
\end{figure}

Waveguiding-based CMOS-MEMS resonators frequently exhibit complicated mode shapes with interesting phononic dispersion behaviour owing to the intricate geometry and range of materials in advanced-node FinFET CMOS technology. A standard abstraction technique is required to reduce the mode shape to an equivalent mechanical description that can be integrated into the overall compact model for the resonator. Theoretical techniques for extracting the equivalent mass, stiffness, and damping of a resonator are not applicable in this scenario. Thus, a technique such as that described in \cite{paramExtraction_COMSOL} can used for CMOS-MEMS RBTs.    

In this paper, a large-signal compact model for a 11.8 GHz Fin Resonant Body Transistor (fRBT) as shown in Fig. \ref{fig11}, fabricated using a commercial 14 nm FinFET (GlobalFoundries\textsuperscript{\textregistered} 14LPP) process has been presented. The organisation of this paper is as follows: Section II gives a detailed description of the unit-cell based 3-D FEM model framework as well as an analysis of the phonon dispersion. In Section III, the equivalent mechanical parameter extraction procedure for the waveguide cavity is described. Section IV addresses the implementation details of the individual constituent modules in the complete fRBT model as well as their interconnections. In Section V, the model is benchmarked against measured resonator data to demonstrate its utility in real-world simulation scenarios. Finally, Section VI presents a conclusion to the study.

\section{Mechanical Resonance and Dispersion Analysis}

A robust, FEM-based, mechanical simulation and analysis framework forms the basis for the compact model of the fRBT. To analyze the mechanical resonance characteristics of the designed fRBT device, a 3D simulation model for a unit cell, as shown in Fig. \ref{fig1}, is constructed in COMSOL Multiphysics\textsuperscript{\textregistered}. Simulations are used to extract the phononic band structure of the acoustic waveguide, which is formed by arraying the fRBT unit cells. The mode of interest and its corresponding dispersion are then analyzed to understand the nature of the displacement and stress fields in the FEOL region. This is relevant for modeling the effect on the drain current as well as equivalent parameter extraction.

\begin{figure}[ht]
\centering
\includegraphics[scale=0.23]{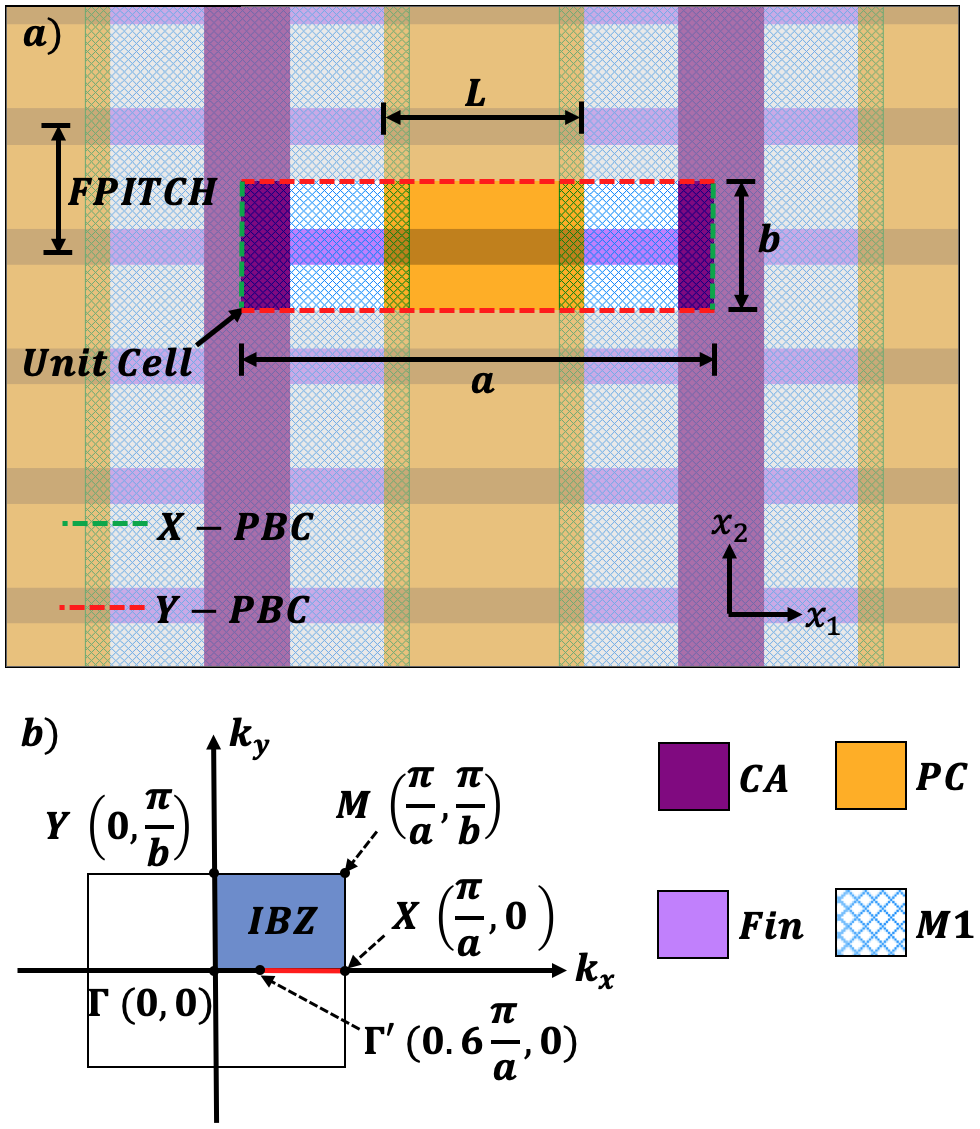}
\vspace{-2mm}
\caption{(a) Cavity region layout including layers up to metal layer M1, highlighting the unit cell along with the X and Y direction PBCs. Higher metal layers and vias are not shown. PC and CA layers in the layout represent the gate and source/drain contacts respectively. (b) Irreducible Brillouin Zone (IBZ) region for the unit cell highlighting symmetry points. Point X is most amenable to excitation with the designed unit cell.}
\label{fig2}
\vspace{-3mm}
\end{figure}

\subsection{Unit Cell 3D FEM Model}

Simulation of an entire 3D model of the resonator is not computationally feasible. Therefore, a pair of repeating unit cells are simulated using periodic boundary conditions (PBCs) along the $x_{1}$ and $x_{2}$ directions. While the use of PBCs to model the resonator cavity introduces a certain degree of inaccuracy because of the finite number of gate interdigitated transducer (IDT) fingers, this is later accounted for in the compact model through an adjustment factor. The FEOL section geometry of the unit cell is constructed using process parameters such as fin height $HFIN$, fin thickness $TFIN$ etc., as well as an understanding of the GF14LPP FinFET process flow. A single gate IDT represents a transistor (number of fingers $NF=1$) with a fin array (number of fins $NFIN$) along the $x_{2}$ direction separated by the technology-defined parameter $FPITCH$ which represents the fin pitch as shown in the layout in Fig. \ref{fig2}(a). The only design parameter available for modification in this section of the geometry is the gate length $L$, which determines the resonance frequency in the particular fRBT mode under consideration in this work. The BEOL portion of the unit cell consists of Mx and Cx level Copper metal phononic crystal (PnC) which is designed such that the phononic bandgap (PnBG) encompasses the resonance frequency to provide confinement along the positive $x_{3}$ direction (out of the plane of the chip). The PnC design takes place within the limits set by the design rules for the process. Each element of the PnC in the BEOL extends uniformly along the gate finger direction $x_{2}$. Appropriate material assignments are done for all regions of the structure to complete the unit cell design. 

Eigenfrequency analysis is required to obtain the mode shapes and corresponding resonance frequencies. Since we only have a single gate IDT per unit cell, the electrical excitation couples most efficiently to the modes corresponding to $k_{x}=\pi/a$ and $k_{y}=0$ ie. point $X$ at the edge of the Irreducible Brillouin Zone (IBZ) as shown in Fig. \ref{fig2}(b). The unit cell is set up to be excited by the wave vector $\overrightarrow{k}=k_{x}\hat{x}_1$ to obtain the eigenstates and the eigenfrequencies where $\hat{x}_1$ represents direction in reciprocal space.

\subsection{Theoretical Formulation for Modal and Dispersion Analysis}
The Plane Wave Expansion framework as described in \cite{PWEbook} can be applied to the phononic waveguide comprising of the fRBT unit cells to obtain a qualitative understanding of the nature of the mode shapes and the dispersion characteristics of the device. The mass density $\rho$ and the elastic moduli $c_{ijkl}$ in the waveguide vary with and are periodic functions of the position vector $\overrightarrow{r}$. If $u_{i}(\overrightarrow{r})$ ($i=1-3$) denotes the displacement field components along the $x_{i}$ directions and $T_{ij}(\overrightarrow{r})$ is the stress, then the Hooke's law can be written in the form:

\begin{equation} \label{eq:HookesLaw}
     T_{ij}(\overrightarrow{r}) = \sum_{kl} c_{ijkl}(\overrightarrow{r}) \frac{\partial u_{k}(\overrightarrow{r})}{\partial x_{l}}
\end{equation}

where, $i$, $j$, $k$ and $l$ can be 1, 2 or 3. The equation of motion in accordance with Newton's second law can we written in the form:
\vspace{-3mm}
\begin{multline} \label{eq:EOM}
    \rho(\overrightarrow{r})\frac{\partial^2 u_{i}(\overrightarrow{r})}{\partial t^2} = \sum_{j} \frac{\partial T_{ij}(\overrightarrow{r})}{\partial x_{j}} \\
    = \sum_{j} \frac{\partial}{\partial x_{j}} \left[ \sum_{kl} c_{ijkl}(\overrightarrow{r}) \frac{\partial u_{k}(\overrightarrow{r})}{\partial x_{l}} \right]
\end{multline}

The materials in the CMOS stack are either isotropic or cubic symmetric in nature which results in the elimination of some of the elements of the $c_{ijkl}$ tensor. Substituting equation (\ref{eq:HookesLaw}) into (\ref{eq:EOM}) and converting to Voigt notation we obtain three coupled equations of motion of the form (position vector dependence of $\rho$ and $c$ has not been shown):

\begin{equation} 
\label{eq:EOM1}
\resizebox{0.45\textwidth}{!}{$%
\begin{aligned}
    \rho \frac{\partial^2 u_{i}}{\partial t^2} = \frac{\partial}{\partial x_{i}} \left( c_{11} \frac{\partial u_{i}}{\partial x_{i}} + c_{12} \left( \frac{\partial u_{j}}{\partial x_{j}} + \frac{\partial u_{k}}{\partial x_{k}} \right)\right) \\
    + \frac{\partial}{\partial x_{j}} \left( c_{44} \left( \frac{\partial u_{i}}{\partial x_{j}} + \frac{\partial u_{j}}{\partial x_{i}} \right) \right) + \frac{\partial}{\partial x_{k}} \left( c_{44} \left( \frac{\partial u_{i}}{\partial x_{k}} + \frac{\partial u_{k}}{\partial x_{i}} \right) \right)
\end{aligned}$%
}    
\end{equation}



where $i$,$j$ and $k$ are 1,2 and 3 respectively for the equation of motion corresponding to the displacement $u_{1}$. Similar equations can be written for the other two components, $u_{2}$ and $u_{3}$. The wave solutions to these equations of motion are of the form $\overrightarrow{u}(\overrightarrow{r})e^{-i\omega t}$ where $\omega$ is the angular frequency. Waves inside a periodic structure such as the fRBT are analogous to plane waves but are modulated by an envelope function. The envelope function takes on the same symmetry and periodicity as the underlying structure. According to the Bloch theorem:

\begin{equation} \label{eq:Bloch}
    \overrightarrow{u}(\overrightarrow{r}) = e^{i\overrightarrow{k}.\overrightarrow{r}} \overrightarrow{U}_{\overrightarrow{k}}(\overrightarrow{r})
\end{equation}

where $\overrightarrow{k}(k_{1}, k_{2}, k_{3})$ is the Bloch wave vector. The term $e^{i\overrightarrow{k}.\overrightarrow{r}}$ in equation \ref{eq:Bloch} represents a plane wave like phase "tilt" term and $\overrightarrow{U}_{\overrightarrow{k}}(\overrightarrow{r})$ is the envelope function. The envelope function has the same periodicity as the fRBT structure such that $\overrightarrow{U}_{\overrightarrow{k}}(\overrightarrow{r})=\overrightarrow{U}_{\overrightarrow{k}}(\overrightarrow{r}+\overrightarrow{R})$. The quantity $\overrightarrow{R}$ belongs to the real space lattice. The envelope function can be expressed in terms of a Fourier series as:

\begin{equation} \label{eq:BlochFourier}
    \overrightarrow{U}_{\overrightarrow{k}}(\overrightarrow{r}) = \sum_{\overrightarrow{G}'}\overrightarrow{U}_{\overrightarrow{k}}(\overrightarrow{G}')e^{i\overrightarrow{G}'.\overrightarrow{r}}
\end{equation}

which leads to

\begin{equation} \label{eq:BlochFourier2}
    \overrightarrow{u}(\overrightarrow{r},t) = e^{-i\omega t} \sum_{\overrightarrow{G}'}\overrightarrow{U}_{\overrightarrow{k}}(\overrightarrow{G}')e^{i(\overrightarrow{G}'+\overrightarrow{k}).\overrightarrow{r}}
\end{equation}

Similarly to the displacement field, $\overrightarrow{u}(\overrightarrow{r})$, $\rho(\overrightarrow{r})$ and $C(\overrightarrow{r})$ can be written in terms of a Fourier expansion using $\alpha(\overrightarrow{r})=(\rho(\overrightarrow{r}),C(\overrightarrow{r}))$:

\begin{equation} \label{eq:BlochFourier3}
    \alpha(\overrightarrow{r}) = \sum_{\overrightarrow{G}''}\alpha(\overrightarrow{G}'')e^{i\overrightarrow{G}''.\overrightarrow{r}}
\end{equation}

where $\overrightarrow{G}'$ and $\overrightarrow{G}''$ belong to the reciprocal lattice domain. Substitution of (\ref{eq:BlochFourier2}) and (\ref{eq:BlochFourier3}) into the equations of the type (\ref{eq:EOM1}), gives the Fourier Transforms of the equations of motion. After some manipulation and modal projection of the resulting equations we get the eigenvalue problem:

\begin{multline} \label{eq:EV1}
    \omega^2 \begin{pmatrix} 
    Q^{(11)}_{\overrightarrow{G},\overrightarrow{G}'} & 0 & 0 \\
    0 & Q^{(22)}_{\overrightarrow{G},\overrightarrow{G}'} & 0 \\
    0 & 0 & Q^{(33)}_{\overrightarrow{G},\overrightarrow{G}'}
    \end{pmatrix}
    \begin{pmatrix}
    U_{1,\overrightarrow{k}}(\overrightarrow{G}') \\
    U_{2,\overrightarrow{k}}(\overrightarrow{G}') \\
    U_{3,\overrightarrow{k}}(\overrightarrow{G}')
    \end{pmatrix} = \\
    \begin{pmatrix}
    P^{(11)}_{\overrightarrow{G},\overrightarrow{G}'} & P^{(12)}_{\overrightarrow{G},\overrightarrow{G}'} & P^{(13)}_{\overrightarrow{G},\overrightarrow{G}'} \\
    P^{(21)}_{\overrightarrow{G},\overrightarrow{G}'} & P^{(22)}_{\overrightarrow{G},\overrightarrow{G}'} & P^{(23)}_{\overrightarrow{G},\overrightarrow{G}'} \\
    P^{(31)}_{\overrightarrow{G},\overrightarrow{G}'} & P^{(32)}_{\overrightarrow{G},\overrightarrow{G}'} & P^{(33)}_{\overrightarrow{G},\overrightarrow{G}'}
    \end{pmatrix}
    \begin{pmatrix}
    U_{1,\overrightarrow{k}}(\overrightarrow{G}') \\
    U_{2,\overrightarrow{k}}(\overrightarrow{G}') \\
    U_{3,\overrightarrow{k}}(\overrightarrow{G}')
    \end{pmatrix}
\end{multline}

which can also be written in the form:

\begin{equation} \label{eq:EV2}
    \omega^2\overleftrightarrow{Q}\overleftrightarrow{U} = \overleftrightarrow{P}\overleftrightarrow{U}
\end{equation}

solution to equation (\ref{eq:EV2}) gives the eigenstates (modes) and their corresponding eigenfrequencies. The elements of the matrices $\overleftrightarrow{P}$ and $\overleftrightarrow{Q}$ determine the polarization and coupling between the modes, respectively. Considering that the fRBT is periodic in the $x_{1}$ and $x_{2}$ directions and the BEOL PnC is also periodic for the metal layers Mx in the positive $x_{3}$ direction, $G_{1}$, $G_{2}$ and $G_{3}$ are assumed to be non-zero. Since we are looking at exciting modes at the symmetry point $X$ using the gate IDT along $x_{2}$, we get $k_{2}, k_{3}=0$. The values of all the elements in the matrices $\overleftrightarrow{P}$ and $\overleftrightarrow{Q}$ can be calculated and it is seen that under the assumptions for the components of $\overrightarrow{G}$ and $\overrightarrow{k}$, none of the terms in the matrix $\overleftrightarrow{P}$ reduce to zero. This means that all of the displacement components can couple to each other in different ways.

\subsection{3-D FEM Modal and Dispersion Analysis}

\begin{figure}\vspace{-2mm}
  \centering
  \begin{tabular}{@{}c@{}}
    \includegraphics[scale=0.24]{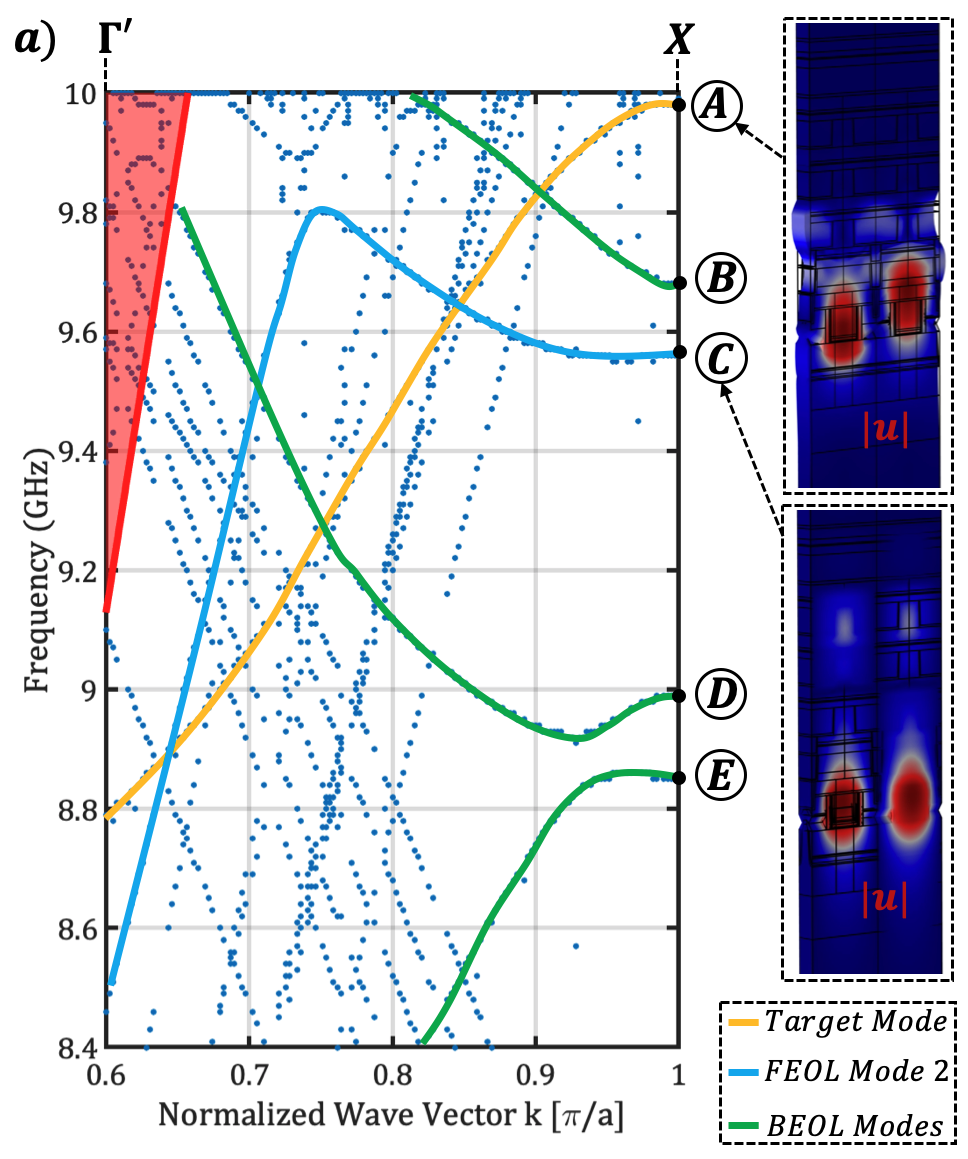} \\
  \end{tabular}


  \begin{tabular}{@{}c@{}}
    \includegraphics[scale=0.24]{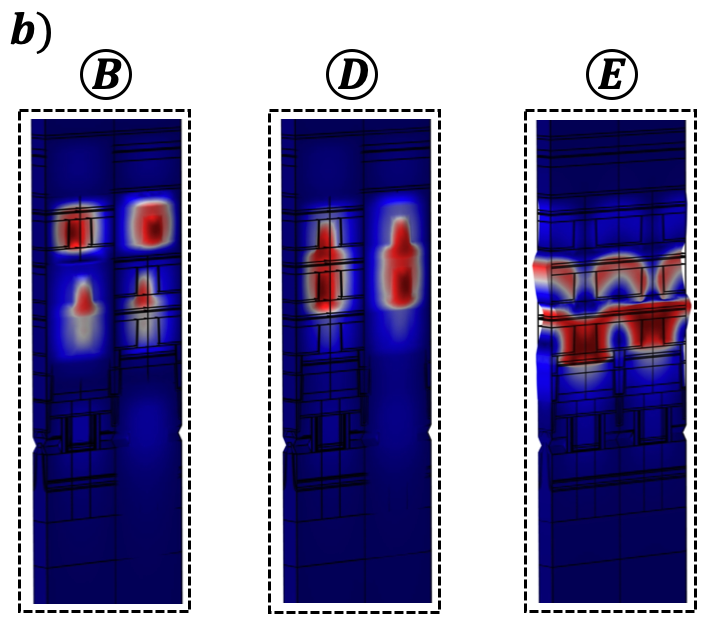} \\
  \end{tabular}

  \caption{(a) Phononic dispersion of the fRBT unit cell obtained using FEM simulation, depicting the targeted mode of operation. (b) Displacement profiles for the PnC modes at $k=1$ (points B, D and E).}\label{fig3}
  \vspace{-6mm}
\end{figure}

The results from the eigenfrequency simulation in COMSOL Multiphysics\textsuperscript{\textregistered} are shown in Fig. \ref{fig3}. As mentioned previously, since the most efficient drive using gate IDTs corresponds to the symmetry point $X$ at the edge of the IBZ, only a part of the $\Gamma -X$ path (highlighted in red in Fig. \ref{fig2}(b)) is traced for the dispersion analysis. The fRBT unit cell is driven with wave vectors with $k_{x}$ ranging from $0.6\pi/a$ to $\pi/a$ and $k_{y}=0$ and the corresponding eigenfrequencies and eigenstates (mode shapes) are evaluated. The resulting dispersion plot as seen in Fig. \ref{fig3}(a), shows the modes lying below the sound cone supported by the structure. The shear wave sound-line in Silicon (represented by the red line $\omega = c_{shear}k_{x}$) demarcates the FEOL-BEOL waveguide mode region from the region where these waveguide modes can couple to bulk modes in the Si substrate. As a general design principle, to reduce scattering to bulk modes, the difference in $\overrightarrow{k}$-space between the mode of interest and the sound-line should be maximized \cite{BahrThesis}. Moreover, for the modes to be sensed and driven efficiently to maximize the electromechanical transconductance $g_{m,em}$, the stress localization should be the strongest in the Si fin of the transistors. Considering these criteria, the mode highlighted in yellow in Fig. \ref{fig3}(a) is the targeted mode for the fRBT. Other modes that are present in Fig. \ref{fig3}(a) are either localized entirely within the BEOL Mx/Cx PnC or Rayleigh modes at the top and bottom extremities of the structure arising due to the finite nature of the simulation model. The modes localised within the BEOL PnC which cannot be driven or sensed efficiently are depicted in Fig. \ref{fig3}(b).

\vspace{2mm}
\section{Equivalent Mechanical Parameter Extraction}

\begin{figure}
\centering
\includegraphics[scale=0.12]{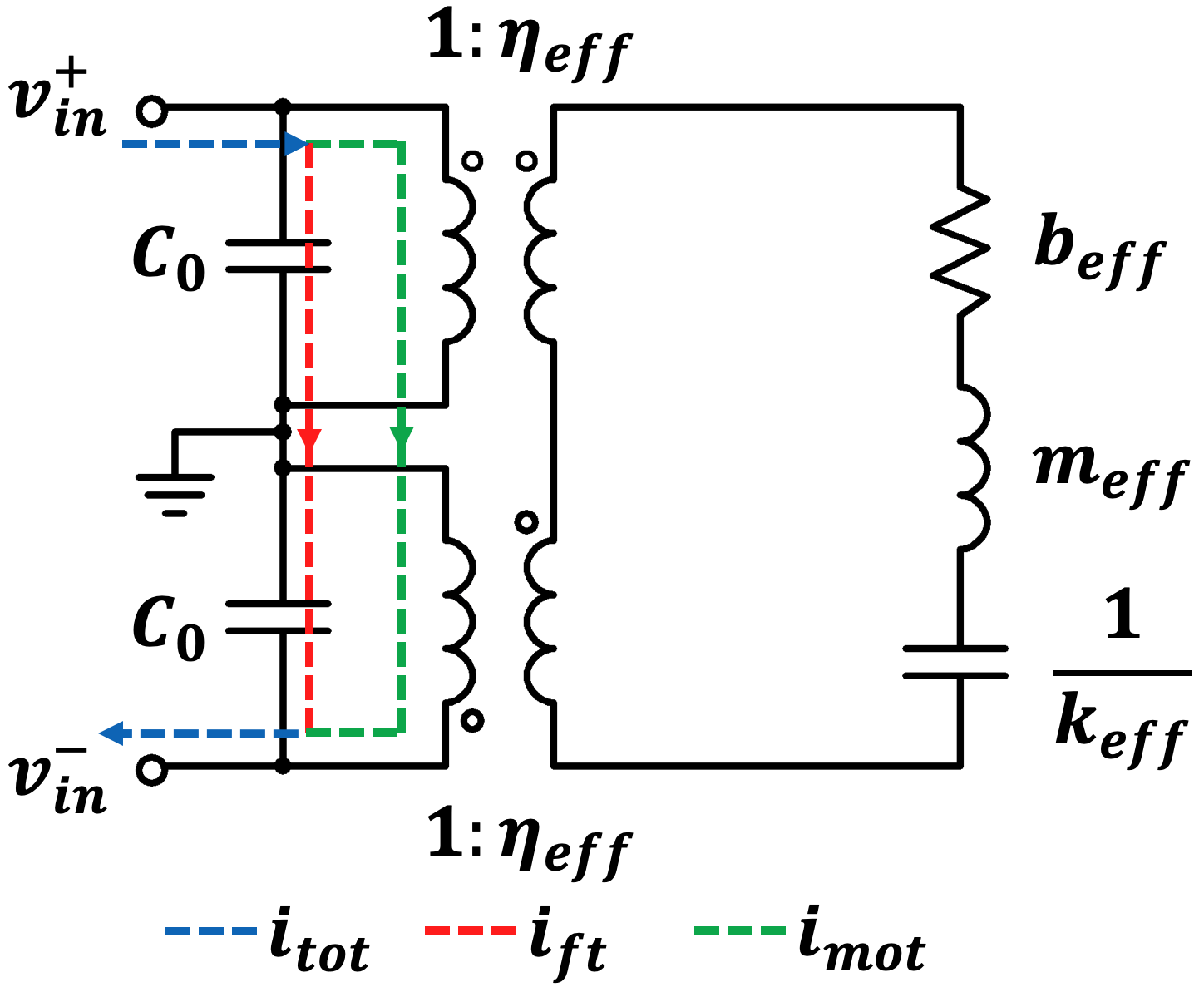}
\caption{Small signal equivalent circuit model of the passive component of the fRBT. The drive MOSCAP capacitance corresponding to each of the two phases in the differential scheme is represented by $c_{0}$, the capacitive electromechanical transduction coefficient by $\eta_{eff}$, and the equivalent mass-spring-damper by $m_{eff}$, $k_{eff}$ and $b_{eff}$, respectively. The total differential current into the differential 1-port $i_{tot}$ is the sum of the feed-through current $i_{ft}$ through $c_{0}$ and the motional current $i_{mot}$.}
\label{fig4}
\end{figure}

Once the target mode for the fRBT has been identified and analyzed, we can then develop its equivalent representation. The 1-D mechanical equivalent for the targeted mode forms the core of the complete fRBT compact model. RF measurements are typically used to extract the equivalent parameters for MEMS resonators. However, this technique is not applicable in the case of fRBT devices operating at frequencies greater than 10 GHz, the reason for which can be explained using the small signal equivalent circuit of the passive section of the device (without the sense transistors) as shown in Fig. \ref{fig4}. At the targeted operation frequencies, the feed-through current $i_{ft}$ through the static MOSCAP capacitance $C_{0}$ is significantly higher than the motional current through the resonator body. This makes the detection of differential 1-port impedance/admittance dependence on frequency exceptionally challenging even with extraction techniques such as that described in \cite{JEYLee_feedthrough}, since the resonance is completely embedded in feed-through.  

An alternative approach for extraction has been described in \cite{paramExtraction_COMSOL} which makes use of eigenfrequency simulations. The main principle behind this approach is the equivalence of the work done in a single degree-of-freedom (DOF) equivalent circuit to that done in a system with multiple DOF such as the fRBT. The transduction scheme in the fRBT is internal dielectric transduction \cite{IDT_JMEMS2009}\cite{RBT_NL2010} via the gate dielectric of the drive MOSCAPs which is different from the externally transduced resonator discussed in \cite{paramExtraction_COMSOL}. Because of a difference in the capacitive transduction mechanism the extraction procedure is required to be amended. The electrostatic work done by the MOSCAP actuator is given by:

\begin{equation} \label{eq:eqParam1}
    W_{e} = |F_{e}|u_{i,eq} = \frac{1}{A_{act}}\iint_{A_{act}}\overrightarrow{F_{e}}.\overrightarrow{u_{i}}dA
\end{equation}

\begin{figure}
\centering
\includegraphics[scale=0.3]{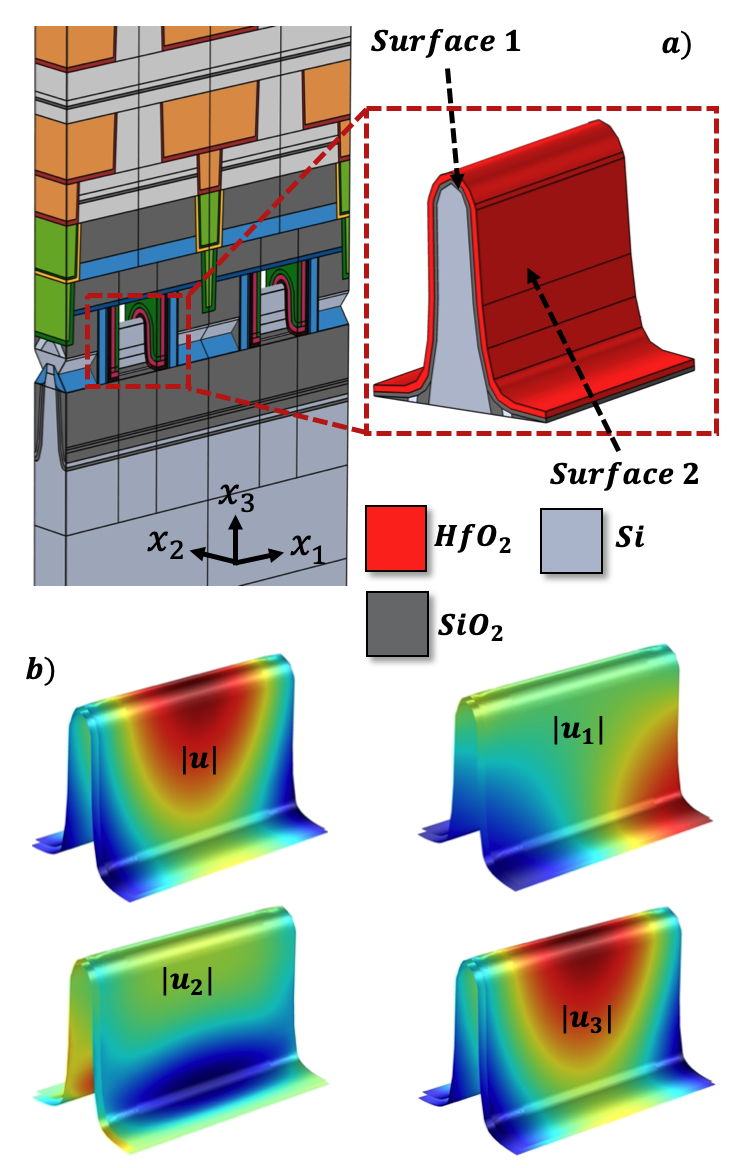}
\vspace{-2mm}
\caption{(a) Fin region of the fRBT showing the actuation surfaces for the electrostatic force. (b) Displacement plots for the actuation surfaces showing the total displacement along with all the components along the three principal directions.}
\label{fig5}
\end{figure}
where $\overrightarrow{F_{e}}$ is the applied electrostatic force, $u_{i,eq}$ the equivalent displacement and $\overrightarrow{u_{i}}$ the vector displacement of the actuation surface having an area $A_{act}$. The electrostatic force is applied between the charge on the metal gate and the charge carriers at the channel surface of the fin. Fig. \ref{fig5}(a) presents the Si fin of the transistor clad with the gate dielectric stack where surface 1 denotes the Si-SiO$_{2}$ interface and surface 2 corresponds to the gate-metal-dielectric interface. Assuming that $A_{1}$ and $A_{2}$ are the areas of surfaces 1 and 2, respectively, the equivalent displacements of the two surfaces after eliminating the force term $|\overrightarrow{F_{e}}|$ in equation \ref{eq:eqParam1} are then given by:


\noindent\begin{minipage}{.49\linewidth}
\begin{equation} \label{eq:eqParam2}
  u_{1} = \frac{\iint_{A_{1}}\overrightarrow{u_{i}}.\hat{n}dA_{1}}{\iint_{A_{1}}dA_{1}}
\end{equation} 
\end{minipage} \vspace{2mm}
\begin{minipage}{.49\linewidth}
\begin{equation} \label{eq:eqParam3}
  u_{2} = \frac{\iint_{A_{2}}\overrightarrow{u_{i}}.\hat{n}dA_{2}}{\iint_{A_{2}}dA_{2}}
\end{equation} 
\end{minipage} \vspace{2mm}

where $\hat{n}$ is the unit vector normal to the actuation surfaces. The gate dielectric is thin and the variation in thickness is negligible as corroborated by the FEM simulations. Thus the equivalent displacement $u_{i,eq}$ can be expressed as an average of the equivalent displacements of the two actuation surfaces:
\begin{equation} \label{eq:eqParam}
    u_{i,eq} = \left| \frac{u_{1}+u_{2}}{2} \right|
\end{equation}

The multi-DOF stored energy in the fRBT unit cell evaluated using a volume integral in COMSOL, $E_{stored}$, is equated to the stored energy in the equivalent single degree of freedom system to give the equivalent parameters:

\noindent\begin{minipage}{.5\linewidth} \vspace{4mm}
\begin{equation} \label{eq:eqParam4}
    k_{eq} = \frac{2E_{stored}}{|u_{i,eq}^2|}
\end{equation}
\end{minipage}%
\begin{minipage}{.5\linewidth} \vspace{4mm}
\begin{equation} \label{eq:eqParam5}
    m_{eq} = \frac{k_{eq}}{|\omega_{0}^2|}
\end{equation}
\end{minipage}
\begin{equation} \label{eq:eqParam6}
    b_{eq} = \frac{\sqrt{m_{eq}k_{eq}}}{Q}
\end{equation}

where $\omega_{0}$ is the resonance frequency and $Q$ is the Q-factor as obtained from the electromechanical transconductance RF measurement. From Fig. \ref{fig5}(b) it can be seen that $|u_{x3}|$ is the dominant component in the total displacement $|u|$. The extracted values of the mechanical parameters for the cavity obtained using FEM eigenfrequency simulation for the targeted mode have been listed in Table I. The same procedure can be used to characterize spurious modes that may be excited besides the main resonance.

\begin{table} \label{table1}
\begin{center}
\footnotesize
\footnotesize{\caption{Extracted Mechanical Parameters}}
\vspace{3mm}
\label{tab_params}
\resizebox{0.65\linewidth}{!}{

    \begin{tabular}{ |c|c| }
    \hline
    PARAMETER & VALUE \\
    \hline
    $m_{eq}$ & $1.0157$ x $10^{-16}$ kg \\
    \hline
    $k_{eq}$ & $0.394$ x $10^{6}$ N/m \\
    \hline
    $b_{eq}$ & $1.054$ x $10^{-8}$ kg/s \\
    \hline
        \end{tabular}
        }
\end{center}
\vspace{-3mm}
\end{table}

\section{Compact Model Implementation}

In this section, based upon the understanding of the fRBT device structure and the targeted mode shape, the implementation details of the constituent building blocks for the complete compact model are discussed. The measured characteristics of the fabricated fRBT device are used for optimizing as well as benchmarking the developed model. The model implementation using VerilogA takes into account compatibility with standard circuit simulators and the PDK for the GF14LPP process.

\subsection{Drive MOSCAP Module}


\begin{figure}[t]
  \centering
  \begin{tabular}{@{}c@{}}
    \includegraphics[scale=0.2]{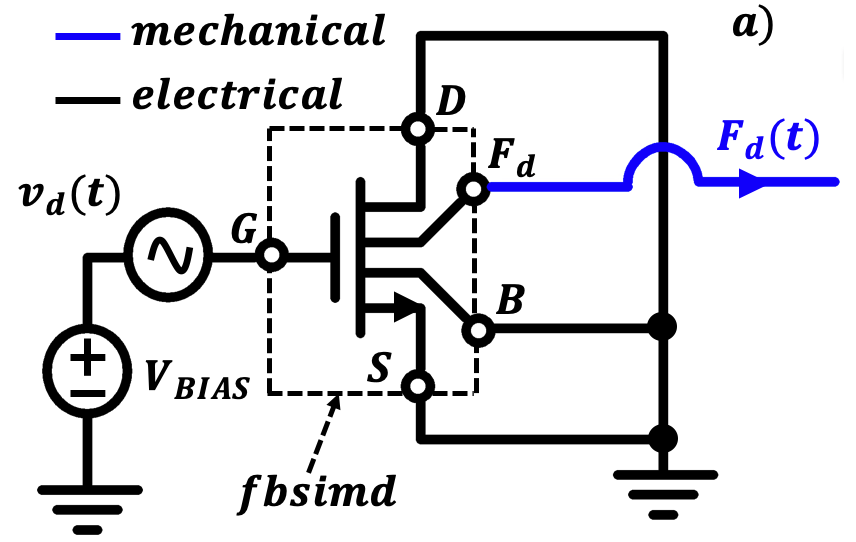} \\
  \end{tabular}


  \begin{tabular}{@{}c@{}}
    \includegraphics[scale=0.23]{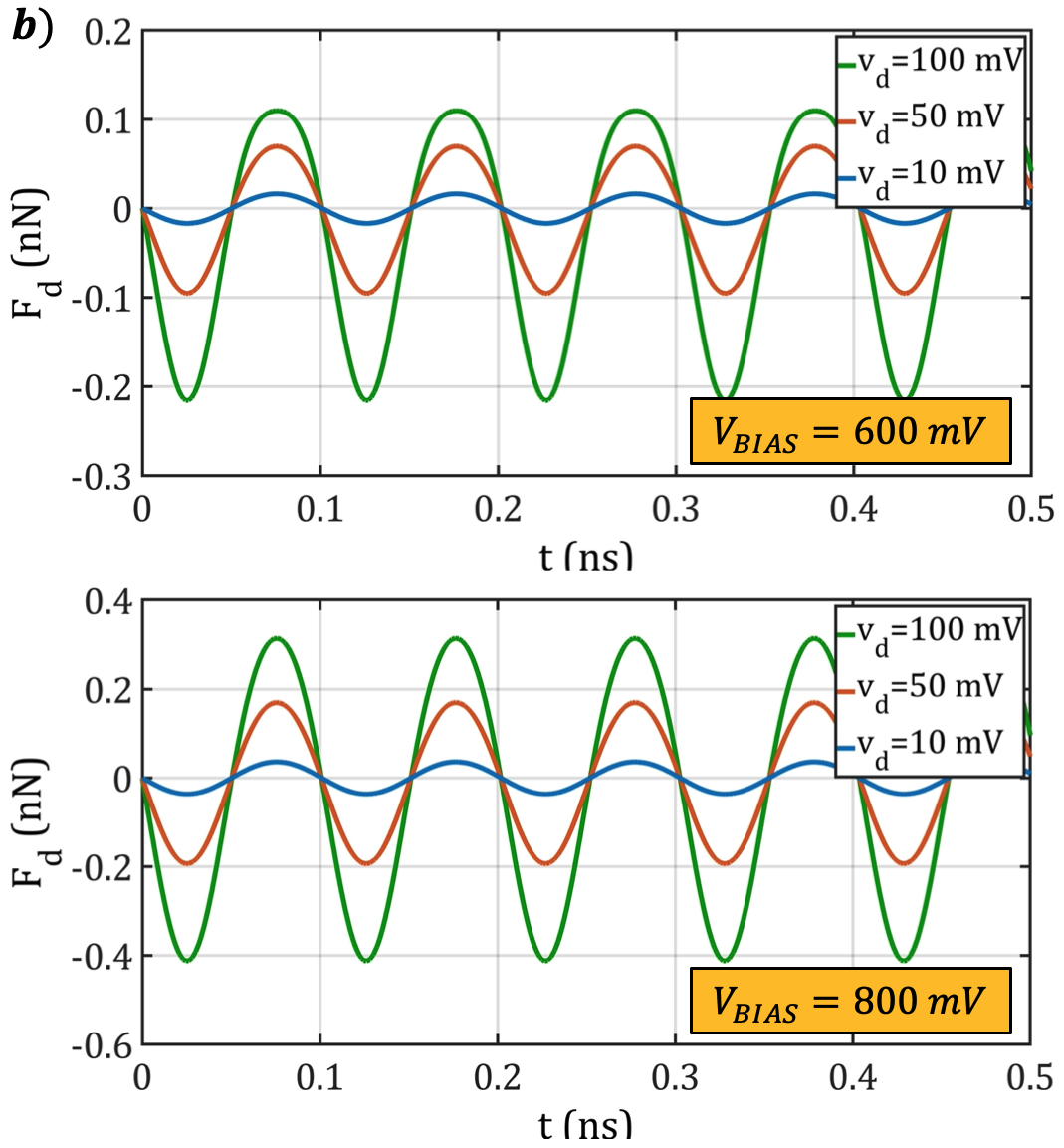} \\
  \end{tabular}

  \caption{(a) Test-bench circuit schematic for the fbsimd transistor. The circuit branch highlighted in blue represents the connection point to the equivalent circuit for the structure's mechanical quantities. (b) Transient electrostatic force output of the fbsimd transistor for three drive voltage levels of 10 mV, 50 mV and 100mV at two gate DC biases of 600 mV and 800 mV.}\label{fig6}
  \vspace{-2mm}
\end{figure}

Standard-V\textsubscript{t} transistors are used to model the array of MOSCAPs that serve as gate-drive transducers. The BSIM-CMG model for FinFET transistors does not account for the electrostatic force generated across the \enquote{plates} of the MOSCAP. Like every other physical capacitor, a force exists between the gate metal and the channel surface charge of the MOSCAP transistor whose source and drain are shorted together. Application of a sinusoidal voltage at the gate results in the generation of an AC force across the gate dielectric which couples to the target mode shape resulting in vibration of the fRBT cavity. The BSIM-CMG model is modified (with the modified model termed \enquote{$fbsimd$}) to include the generation of this electrostatic actuation force and a corresponding additional mechanical terminal is added alongside the original gate, drain, source, and bulk electrical terminals. 

From \cite{BSIMCMGman} it can be seen that the standard BSIM-CMG model calculates intrinsic capacitances of a transistor using the terminal charges at all four terminals. Similarly, in $fbsimd$, the calculated gate charge $Q_{G}$ and intrinsic gate capacitance $c_{ox}$ are used to evaluate the driving voltage $v_{D}$ across the gate dielectric film:

\begin{equation} \label{eq:vd}
    v_{D}(t) = \frac{Q_{G}(t)}{c_{ox}(t)}
\end{equation}
 
The drive force generated by a drive MOSCAP IDT is then given by:

\begin{equation} \label{eq:Fd}
    F_{d}(t) = -\chi\frac{\epsilon_{0}\epsilon_{r}}{2}\frac{Av_{D}^2(t)}{t_{ox}^2}
\end{equation}

where A is the effective actuation area of the drive transducer that depends on the number of fins $NFIN$, other fin geometry parameters as well as the gate length $L$, $t_{ox}$ is a model parameter representing the gate dielectric thickness and $\epsilon_{r}$ is the relative gate dielectric stack permittivity. The model parameter $\chi$ is a drive force adjustment parameter that is introduced to account for the effect of a finite transducer length which modifies effective coupling to the cavity mode shape.    

Transient simulation for the $fbsimd$ transistor is carried out using the test-bench as shown in Fig. \ref{fig6}(a) and the resulting force waveforms are analysed. As can be seen in each of the DC biasing cases of Fig. \ref{fig6}(b), for small drive voltage levels such as 10 mV, the distortion of the output force waveform $F_{d}(t)$ is minimal. However, for large drive voltage levels such as 100 mV, the distortion in $F_{d}(t)$ is appreciable. The nonlinearity in the $F_{d}(t)$ waveform can be attributed to the bias dependent nature of the MOSCAP capacitance which fluctuates when the $fbsimd$ transistor gate is driven by a large signal. The distortion increases as the DC biasing reaches closer to the $V_{t}$ of the $fbsimd$ (which is same as the standard-$V_{t}$ n-channel FinFET). The drive MOSCAPs should therefore be operated in the strong inversion or accumulation regimes. 

\subsection{Resonant Body Module}

Using the equivalent mechanical parameters for the fRBT waveguide cavity extracted in Section III, a mechanical resonant body module has been constructed. Spurious mode resonances are added alongside the target resonance, all of which can be modeled using the force-current formulation \cite{MEMS_Senturia}. The state-space representation for the $i^{th}$ mode in the module is as follows:

\begin{subequations} \label{eq:resBodyEq}
    \begin{equation}
        \nu_{i}(t) = \frac{dx_{i}(t)}{dt}
    \end{equation}
    \begin{equation}\vspace{2mm}
        F_{d,i}(t) = m_{eq,i}\frac{d\nu_{i}(t)}{dt}+b_{eq,i}\nu_{i}(t)+k_{eq,i}x_{i}(t)+\sqrt{4k_{B}Tb_{eq,i}}
    \end{equation}
    \begin{equation} \vspace{2mm}
        F_{d,i}(t) = \gamma_{i}F_{d}(t)
    \end{equation}
\end{subequations}

where $v_{i}(t)$ is the velocity associated with the mechanical node $x_{i}(t)$, $k_{B}$ is the Boltzmann Constant and $T$ is the temperature. The $\sqrt{4k_{B}Tb_{eq,i}}$ term added to the model to account for the mechanical force noise in the device. The term $\gamma_{i}$ is used to model the efficiency with which the generated electrostatic force $F_{d}(t)$ couples to the $i^{th}$ mode. The force-current analogy is best suited for implementation of the resonant body since it allows the addition of multiple drive modules, each of which contributes a drive force current. To combine the effects of all the modes considered in the model, the velocities corresponding to each mode are summed:

\begin{equation}
    \nu(t) = \sum_{i=1}^{N} \nu_{i}(t)
\end{equation}

Typically, series $RLC$ branches are added in parallel corresponding to each mode as shown in \cite{multiModeMBVD}. The complement of the this circuit convention is considered in the case of the resonant body module of the fRBT since the force-current analogy is employed.

\subsection{Sense Transistor Module}

Although the drive MOSCAP transducers together with the resonant cavity are sufficient to create a resonator, owing to the difficulties with RF detection of the resonator response in an all-passive implementation as highlighted in section III, transistor-based readout is the optimal choice for the fRBT device. Two standard-$V_{t}$ transistors which act as vibration sensors are embedded in the centre of the waveguide cavity. These transistors are exactly the same as the drive MOSCAP transistors except that they are connected and biased so as to conduct current through their channels. The time-varying stress in the waveguide cavity causes a modulation in the drain current of the sense transistors which results in a differential current readout. Stress induced in the channel of the sense FinFET transistors causes changes to the electronic band structure of the channel material (in this case Silicon). This effect primarily manifests itself in the form of variation in carrier mobility $\mu_{n}$ through the piezoresistive effect, threshold voltage $V_{t}$, and saturation velocity $\nu_{sat}$.

The effect of stress on transistor characteristics has been studied extensively in literature in the case of Layout Dependent Effects (LDEs) due to shallow trench isolation (STI) etc. \cite{stressEffect1}\cite{stressEffect2}\cite{stressEffect3} and strained-silicon transistors for mobility enhancement \cite{stressEffect4}. The parameters used for modeling the dependence of drain current on stress in the conventional BSIM-CMG model become time-variant in the case of the fRBT. A new modified BSIM-CMG model called \enquote{$fbsims$} is therefore developed with an additional mechanical terminal (similar to the $fbsimd$ model) and which can calculate the time-varying changes to $\mu_{n}$, $V_{t}$ and $\nu_{sat}$. 

\begin{figure}[ht]
  \centering
  \begin{tabular}{@{}c@{}}
    \includegraphics[scale=0.13]{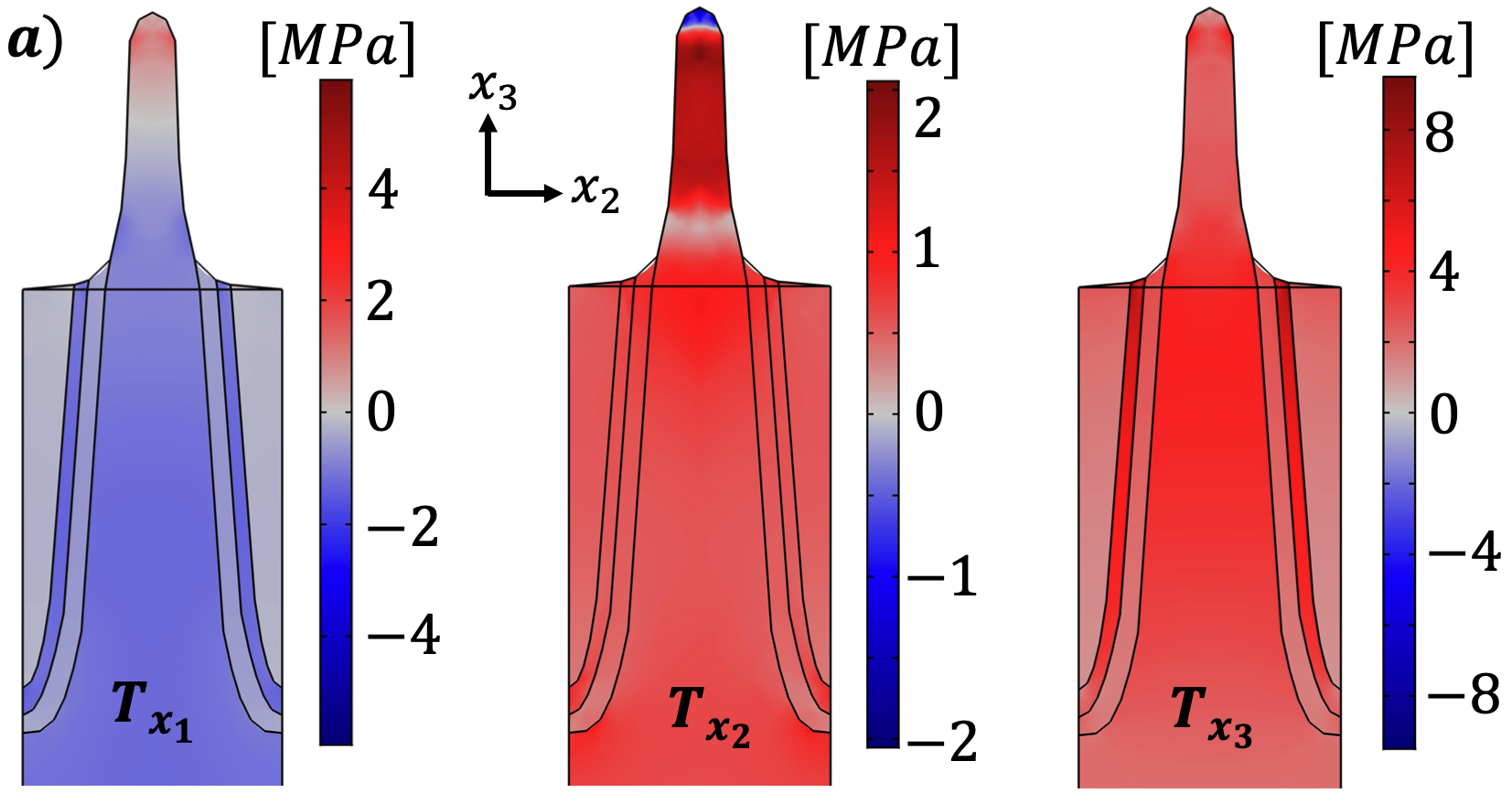} \\
  \end{tabular}


  \begin{tabular}{@{}c@{}}
    \includegraphics[scale=0.13]{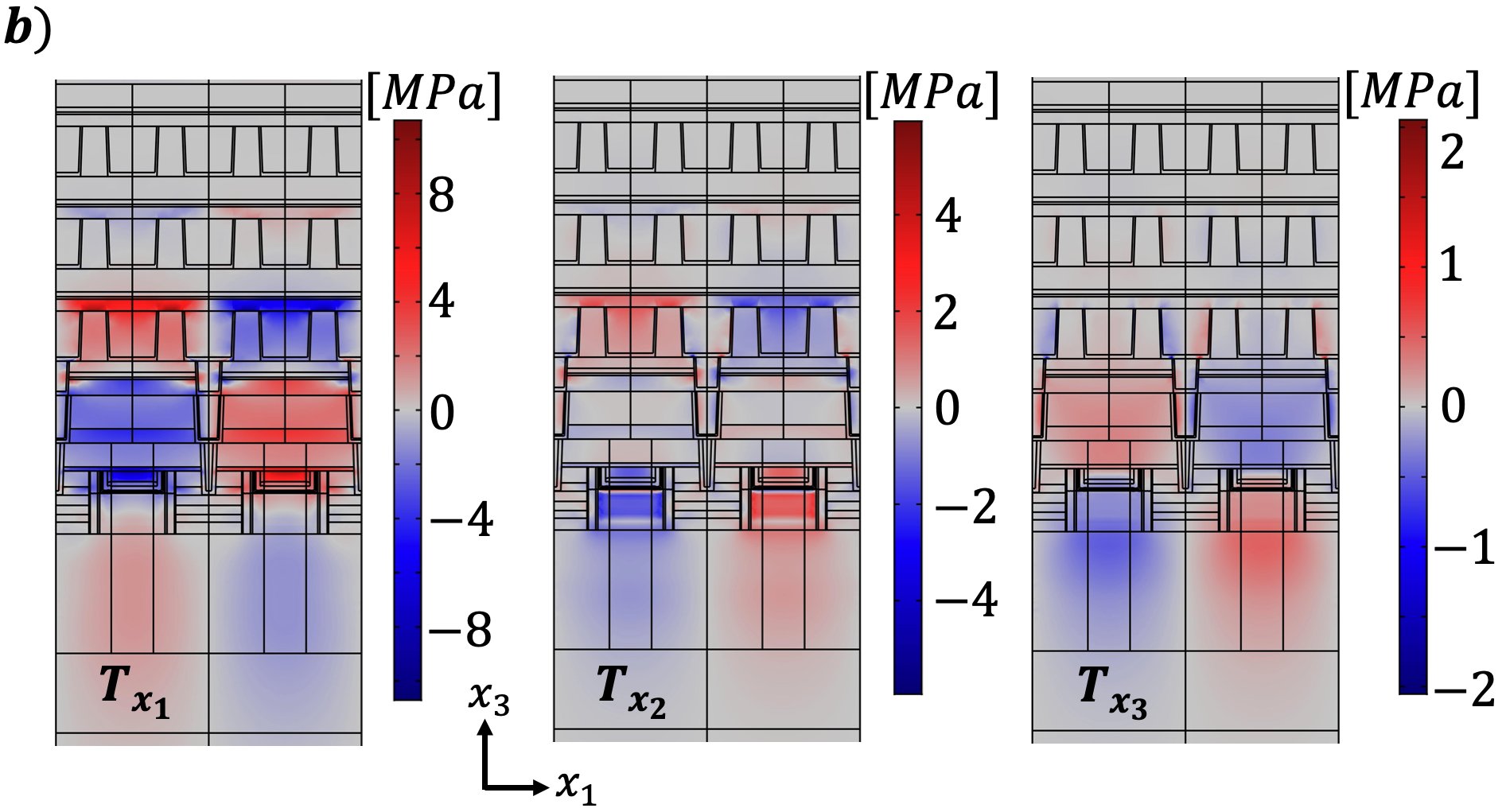} \\
  \end{tabular}

  \caption{(a) Stress profiles of the target mode shape for the $x_{2}-x_{3}$ plane in the fin region. (b) Stress component profiles for a pair of adjacent unit cell gate transducers in the $x_{1}-x_{3}$ plane, depicting energy localization in the FEOL region. Opposite signs for the stress are seen in the adjacent transducers because of the differential nature of the mode shape. A drive stress of 6 MPa has been considered for the frequency domain simulation.}\label{fig7}
  \vspace{4mm}
\end{figure}

It is important to analyze the nature of time varying stresses and strains in the sense transistor fin region before their effect on transistor properties can be modeled. In Fig. \ref{fig7}(a) we can see the stress profiles in the fin region along the $x_{2}-x_{3}$ plane corresponding to the target mode shape. Considering the current density for the "on" state in a tapered fin as simulated in \cite{FinFETcurrDen}, it can be noted that the distribution is uniform across the majority of the fin cross-section in the $x_{2}-x_{3}$ plane. Moreover, as seen in Fig. \ref{fig7}(b), all three stress components are also uniform in the fin region along the $x_{1}$ direction. Therefore, we can use a volume averaged stress formulation in the fin region to calculate stress values for transistor parameter variations:

\begin{equation} \label{eq:sigma}
    \overline{T} = \frac{\iiint_V T dx_{1}dx_{2}dx_{3}}{\iiint_V dx_{1}dx_{2}dx_{3}}
\end{equation}
\vspace{2mm}
\subsubsection{Mobility Modulation Model}

\begin{table} \label{table2}
\vspace{-6mm}
\begin{center}
\footnotesize
\footnotesize{\caption{FINFET PIEZORESISTIVITY COEFFICIENTS \cite{FinFET_Piezo_Coeff}}}
\label{tab_params3}
\resizebox{\linewidth}{!}{

    \begin{tabular}{ |c|c|c|c|}
    \hline
      & $\pi_{11}'$ (Pa\textsuperscript{-1}) & \textbf{$\pi_{12}'$} (Pa\textsuperscript{-1}) & \textbf{$\pi_{12}$} (Pa\textsuperscript{-1}) \\
    \hline
     NFET & $-45.2$x$10^{-11}$ & $-25.6$x$10^{-11}$ & $57.6$x$10^{-11}$ \\
    \hline
     PFET & $45.0$x$10^{-11}$ & $-23.8$x$10^{-11}$ & $-10.1$x$10^{-11}$ \\
    \hline
        \end{tabular}
        }
\end{center}
\vspace{-7mm}
\end{table}

Piezoresistivity theory for Si can be used to model the effect of channel stress on the electron mobility $\mu_{n}$. It is assumed that the FinFET transistor channel in the GF14LPP process is oriented along the $\langle 110 \rangle$ direction so as to boost the weaker hole mobility $\mu_p$. Thus, the piezoresistivity coefficients of Silicon as described in \cite{Kanda} and \cite{Jaeger}, which are based on the $\langle 100 \rangle$ crystallographic axis, must be modified to account for the coordinate system rotation by 45\textsuperscript{o}. Also, the piezoresistive coefficients in Si FinFET devices differ from those in bulk Si \cite{FinFET_Piezo_Coeff} and hence, the values (listed in Table II) for the $\langle 110 \rangle$ direction are used in this work. The relative change in mobility with stress is given by the relation \cite{SravanThesis}:

\begin{equation} \label{eq:dmumu}
    \frac{\Delta\mu}{\mu} = \pi_{11}'T_{x_{1}x_{1}} + \pi_{12}'T_{x_{2}x_{2}} + \pi_{12}T_{x_{3}x_{3}}
\end{equation}

where
\begin{subequations} \label{eq:pidash}
    \begin{equation} \label{eq:pi11dash}
        \pi_{11}' = \frac{\pi_{11}+\pi_{12}+\pi_{44}}{2}
    \end{equation}
    \begin{equation} \label{eq:pi12dash}
        \pi_{12}' = \frac{\pi_{11}+\pi_{12}-\pi_{44}}{2}
    \end{equation}
\end{subequations}
In equation (\ref{eq:dmumu}), $T_{x_{1}x_{1}}$, $T_{x_{2}x_{2}}$ and $T_{x_{3}x_{3}}$ represent the time-varying, volume-averaged stresses obtained using equation (\ref{eq:sigma}) from COMSOL. While $\pi_{11}$, $\pi_{12}$ and $\pi_{44}$ represent the piezoresistivity coefficients along the $\langle 100 \rangle$ crystallographic axes, $\pi_{11}'$ and $\pi_{12}'$ represent piezoresistivity coefficients for the rotated system along the $\langle 110 \rangle$ directions corresponding to the $x_{1}$, $x_{2}$ and $x_{3}$ directions. 

A mobility multiplier $1+\frac{\Delta\mu}{\mu}$ is used in conjunction with the mobility degradation factor $D_{mob}$ \cite{BSIMCMGman} in the drain current equation of the $fbsims$ model to account for the variation of the transistor drain current with the stress in the sense transistor fin.

\vspace{3mm}
\subsubsection{$V_{t}$ Modulation Model}

Stresses induced in the channel region cause changes to the band structure, which result in fluctuations in the band-edge potentials, band-gap, and the effective density of states \cite{stressEffect1}. Due to these shifts, the flatband voltage $V_{FB}$ and channel surface potential $\psi_{s}$ change, causing a change in the $V_{t}$. Threshold voltage is typically treated as a static parameter in the BSIM-CMG model with variability parameters such as $DELVTRAND$ \cite{BSIMCMGman} introduced to model the effect of layout etc. on the $V_{t}$. In this work, the model used in \cite{Sravan_FinFETstress} is adapted for calculating the shift in conduction band-edge potential $\Delta E_{c}$ due to time-varying strains in the channel. Changes to the valence band-edge potential $\Delta E_{v}$ are evaluated using the model in \cite{stressEffect1}. Both conduction and valence band-edge potential shifts are evaluated as:

\begin{subequations} \label{eq:bandEdge}
    \begin{equation}
        \Delta E_{c} = \Xi_{d}(S_{x_{1}x_{1}} + S_{x_{2}x_{2}} + S_{x_{3}x_{3}} ) + \Xi_{u} S_{x_{3}x_{3}}
    \end{equation}
    \begin{equation}
        \Delta E_{v} = u_{1}(S_{x_{1}x_{1}} + S_{x_{2}x_{2}} + S_{x_{3}x_{3}} ) + 2u_{2}(S_{x_{3}x_{3}} - S_{x_{1}x_{1}})
    \end{equation}
\end{subequations}

\begin{figure}[t]
  \centering
  \begin{tabular}{@{}c@{}}
    \includegraphics[scale=0.11]{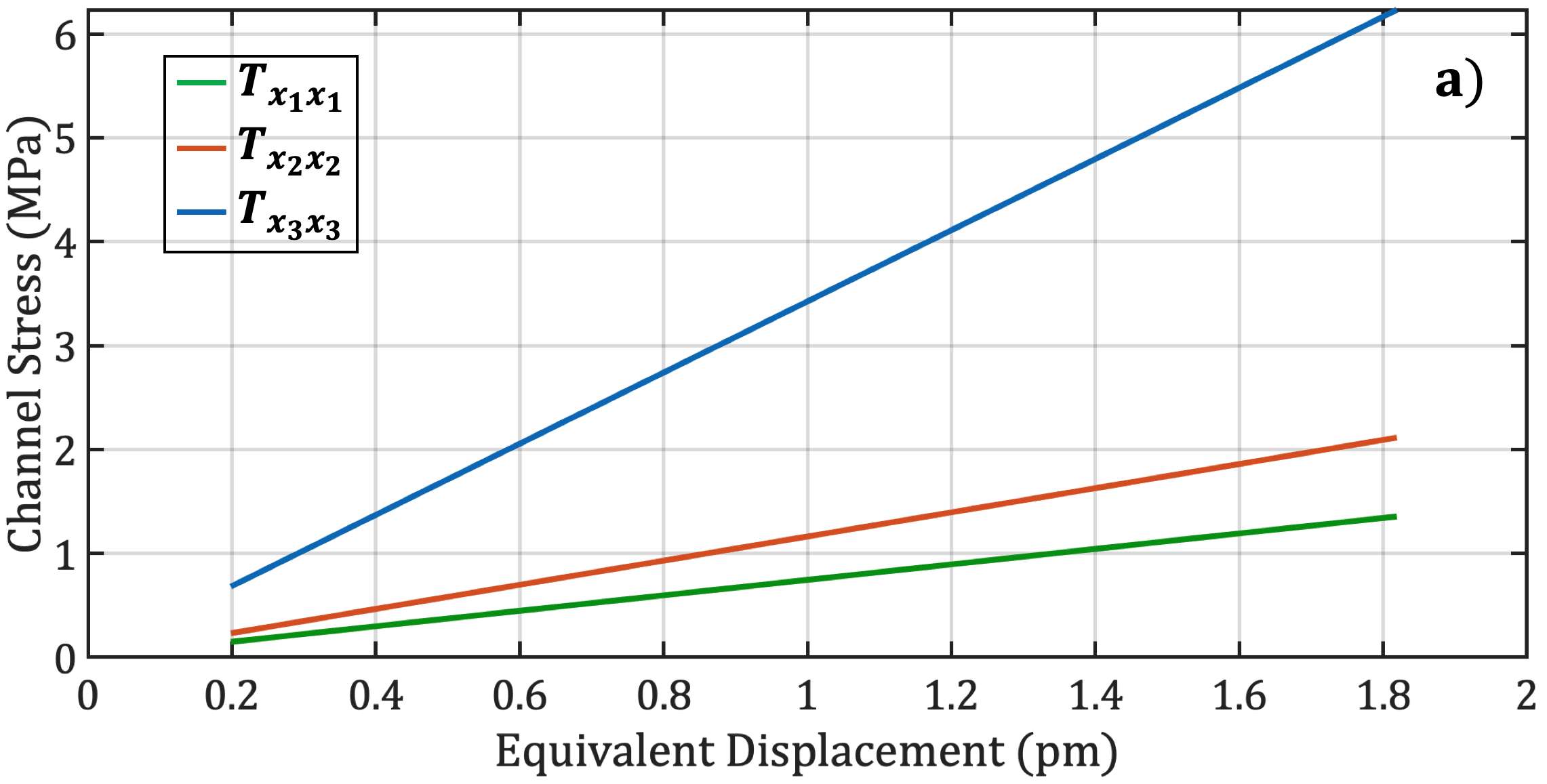} \\
  \end{tabular}


  \begin{tabular}{@{}c@{}}
    \includegraphics[scale=0.108]{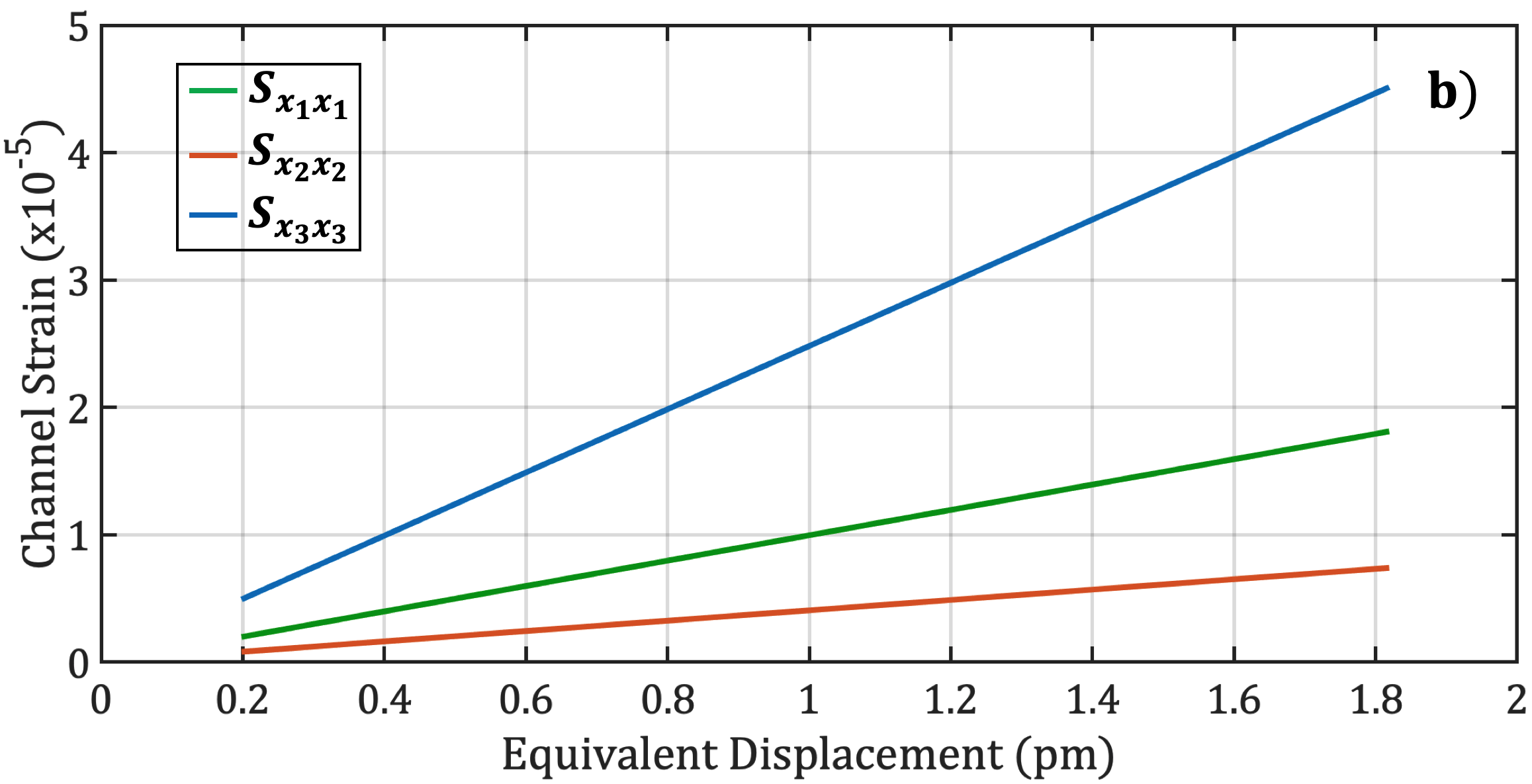} \\
  \end{tabular}

  \caption{(a) Plot of channel volume-averaged stress versus the equivalent displacement as evaluated from frequency domain simulation in COMSOL (b) Plot of channel volume-averaged strain versus the equivalent displacement.}\label{fig8}
  \vspace{-2mm}
\end{figure}
while $S_{x_{1}x_{1}}$, $S_{x_{2}x_{2}}$ and $S_{x_{3}x_{3}}$ are time-varying, channel-averaged strain components, $\Xi_{d}$, $\Xi_{u}$, $u_{1}$ and $u_{2}$ are deformation potential constants (values given in Table III). The shifts in the threshold voltage in the n- or p-type transistors are given by:

\begin{subequations} \label{eq:VtShift}
    \begin{equation}
        q\Delta V_{thp} = (m-1)\Delta E_{c} - m\Delta E_{v}
    \end{equation}
    \begin{equation}
        q\Delta V_{thn} = -m\Delta E_{c} + (m-1)\Delta E_{v}
    \end{equation}
\end{subequations}

where $m$ is the body effect parameter. This shift in the $V_{t}$ is incorporated in the $fbsims$ model for the sense transistor alongside the variability parameter $DELVTRAND$.
\vspace{3mm}
\begin{table} \label{table3}
\begin{center}
\footnotesize
\footnotesize{\caption{THRESHOLD VOLTAGE MODEL CONSTANTS \cite{stressEffect1}\cite{Sravan_FinFETstress}}}
\label{tab_params2}
\resizebox{0.6\linewidth}{!}{
    \begin{tabular}{|c|c|c|}
    \hline
     PARAMETER & VALUE & UNIT \\
    \hline
        $\Xi_{d}$ & $1.13$ & $eV$ \\
    \hline
        $\Xi_{u}$ & $9.16$ & $eV$ \\
    \hline
        $u_{1}$ & $2.46$ & $eV$ \\
    \hline
        $u_{2}$ & $-2.35$ & $eV$ \\
    \hline
         $m$ & $1.3$ & $1$ \\
    \hline
    \end{tabular}
        }
\end{center}
\vspace{-4mm}
\end{table}

\begin{figure}[t]
  \centering
  \begin{tabular}{@{}c@{}}
    \includegraphics[scale=0.17]{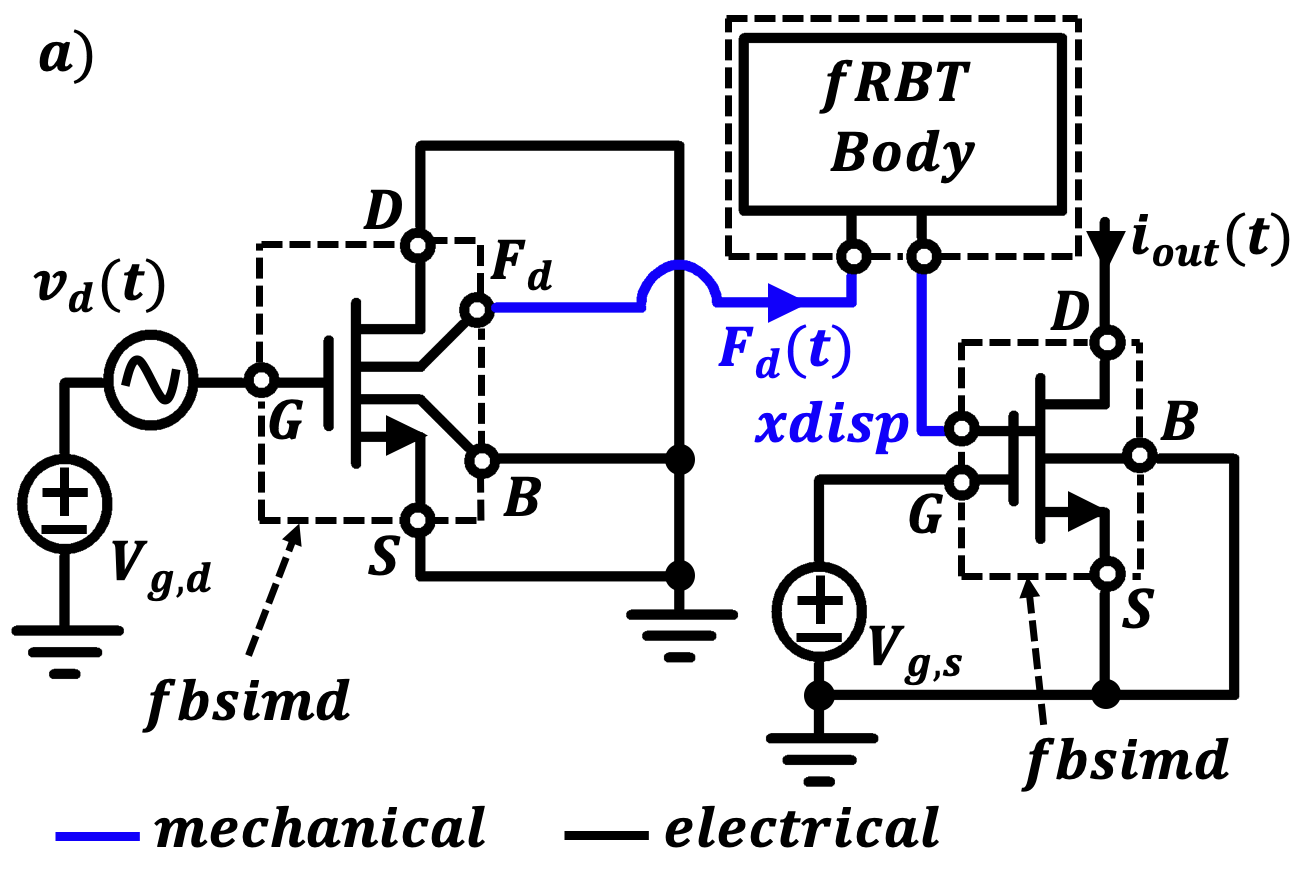} \\
  \end{tabular}

  \begin{tabular}{@{}c@{}} \hspace{-10mm}
    \includegraphics[scale=0.12]{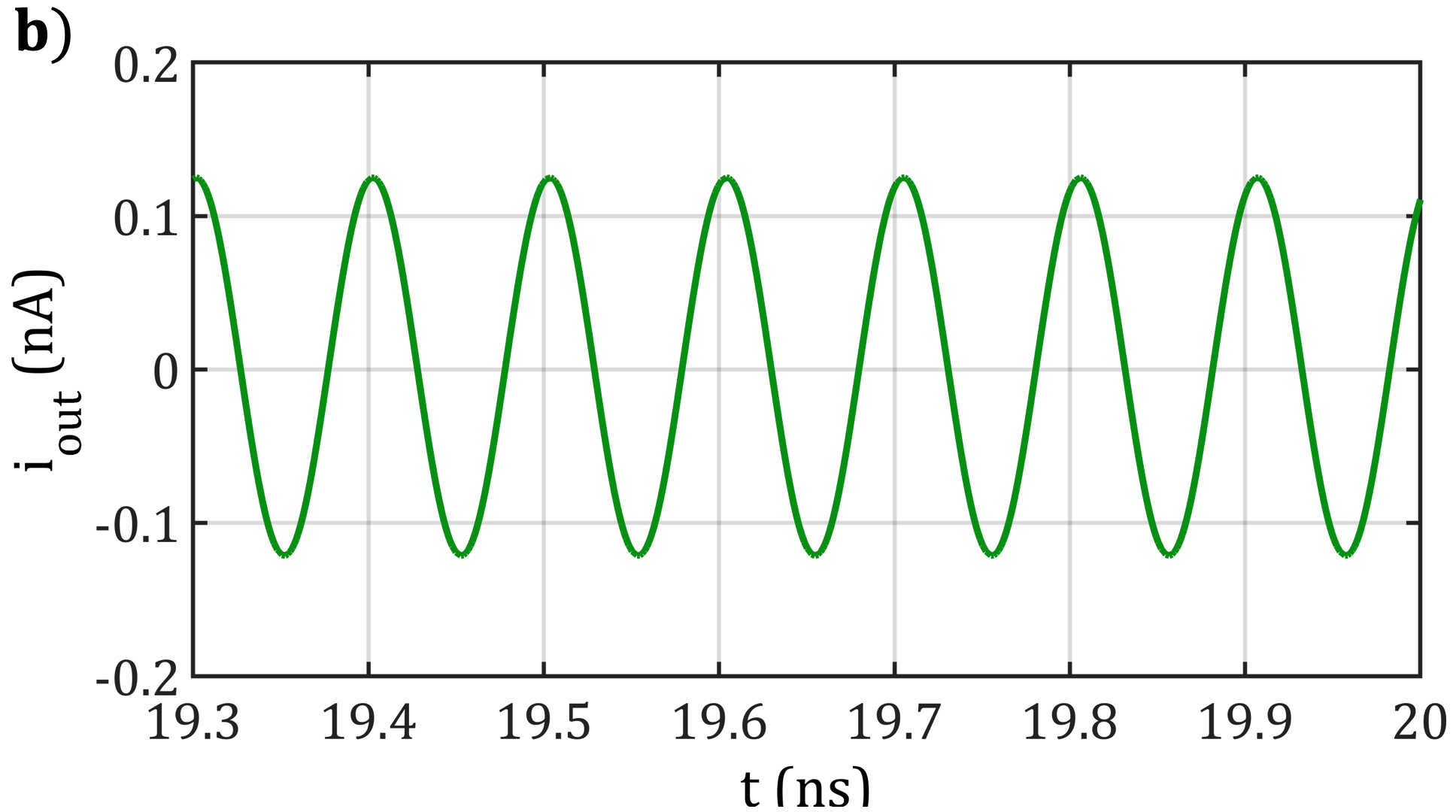} \\
  \end{tabular}

  \begin{tabular}{@{}c@{}}
    \includegraphics[scale=0.105]{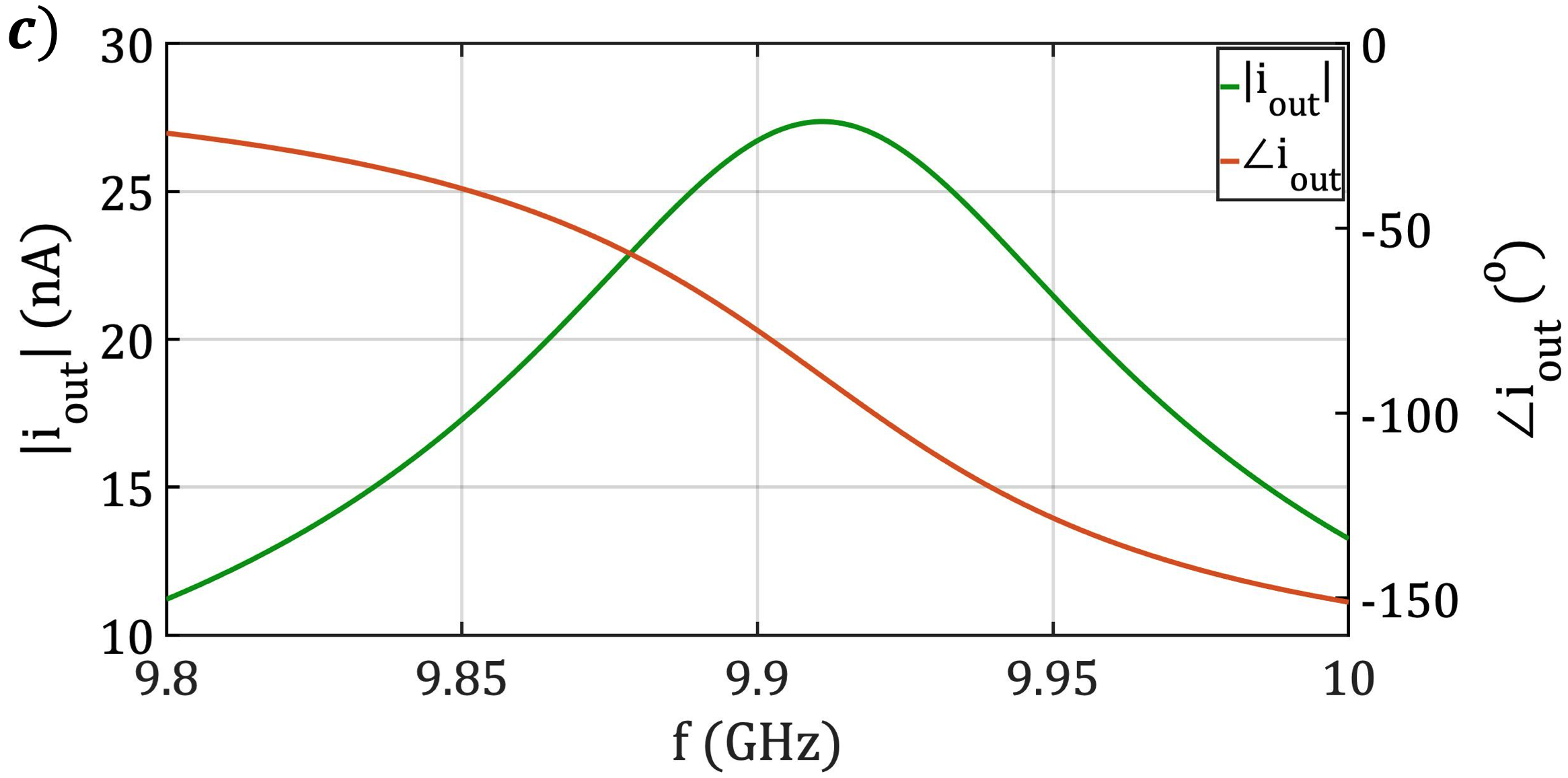} \\
  \end{tabular}

  \caption{(a) Testbench circuit schematic for evaluating the performance of the fbsims sense transistor. The fbsims is used in conjunction with the fbsimd and fRBT resonant body module. (b) Transient current output of the fbsims transistor with a 10 mV drive at the fbsimd MOSCAP input. (c) Magnitude and phase plots for the output current depicting a resonance at the frequency set by the fRBT resonant body module.}\label{fig9}
  \vspace{-2mm}
\end{figure}

\subsubsection{$\nu_{vsat}$ Modulation Model}

The effect of stress on the saturation velocity $\nu_{sat}$ can be modeled using the methodology described in \cite{stressEffect1} and \cite{stressEffect5}. There is a direct relationship between the $\nu_{sat}$ variation in highly scaled devices with ballistic efficiency approaching 1 and variation in mobility due to stress, through a factor $\alpha$. The $\nu_{sat}$ modulation is then given by:

\begin{figure*}[t]
\centering
\includegraphics[scale=0.225]{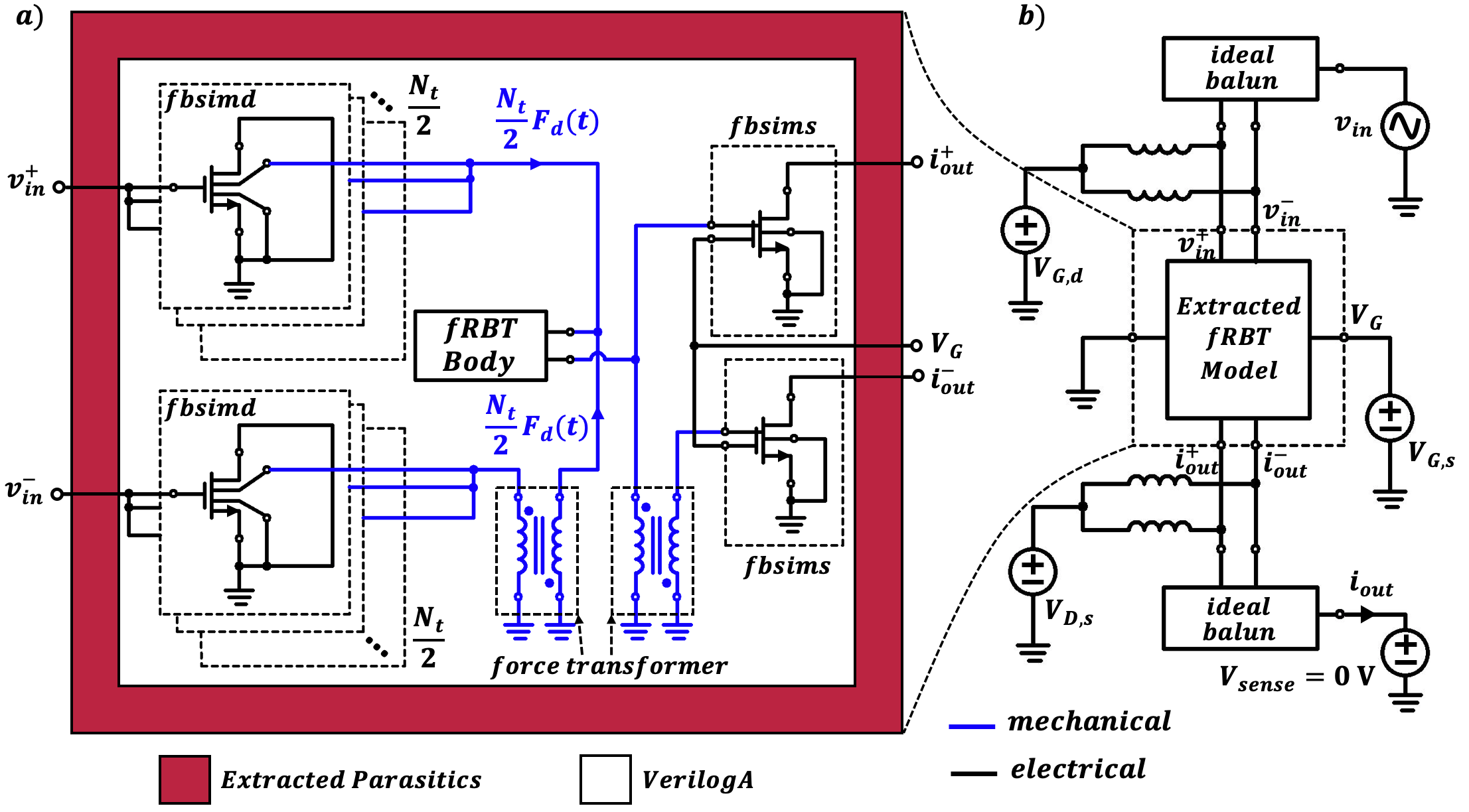}
\vspace{-2mm}
\caption{(a) Complete model for the fRBT device comprising of the VerilogA building blocks as well as parasitics extracted from the device layout. (b) Testbench mimicking the RF measurement setup for testing the characteristics of the extracted fRBT model.}
\label{fig10}
\end{figure*}

\begin{equation} \label{eq:vsatMod}
        \nu_{sat} = \nu_{sat0} \left(1 + \alpha \frac{\Delta\mu}{\mu}\right)
\end{equation}

where $\nu_{sat0}$ is the static saturation velocity and the constant $\alpha$ has a value of $0.85$. Source-drain current degradation due to $\nu_{sat}$ is modeled in the BSIM-CMG model using the degradation parameter $D_{vsat}$ \cite{BSIMCMGman} so the $\nu_{sat}$ modulation is added in conjunction to this parameter in the $fbsims$ model.
\vspace{3mm}

The sense transistor module employing the $fbsims$ model uses stress and strain values to calculate the changes in $\mu$, $V_{t}$ and $\nu_{sat}$. The resonant behaviour of the device however, is captured using equivalent displacement parameter through the resonant body module as described in Section IV-B. Therefore, the equivalent displacement must be converted to stress and strain components which can then be utilized. To evaluate the relationship between the equivalent displacement and the stress/strain in the channel, frequency domain simulation was carried out in COMSOL at the resonance frequency while varying the drive stress level. The linear relationship between the equivalent displacement and the channel stress/strain is presented in Fig. \ref{fig8}. The constants of proportionality obtained from this analysis are used to perform the equivalent displacement to stress/strain conversion in the $fbsims$ model. We implement a testbench shown in Fig. \ref{fig9} to obtain transient and frequency sweep characteristics of the output current of a $fbsims$ transistor. The $fbsimd$ drive MOSCAP and fRBT body modules are added, and the mechanical terminals for all three components are connected to the same mechanical node. When both the $fbsims$ and $fbsimd$ are biased in strong inversion and a drive voltage of 10 mV is applied to the input, a transient output waveform is observed as shown in Fig. \ref{fig9}(b). Under the same biasing conditions, an ac simulation is performed to obtain the frequency response of the "fbsims" transistor. Fig. \ref{fig9}(c) shows the resonant characteristic of the output current, as expected from the model when used with the drive and body modules.

Once the individual building blocks of the fRBT model have been implemented and tested for correct functionality, they are connected together in the same configuration as the actual device as shown in Fig. \ref{fig10}(a). The differential drive section of the model consists of $N_{t}$ $fbsimd$ drive transducers, half of which are connected to one phase of the input drive voltage and half to the opposite phase. The force contribution from the drive units is then added to obtain the total drive force for the fRBT body module. To ensure that the force from the differential drive transducers add in phase, 1:1 mechanical transformers are implemented which invert the polarity of the mechanical quantities being carried by the mechanical network. The mechanical node $x_{disp}$ is common to all the modules in the fRBT model. The same mechanical transformer is also used in conjunction with one of the $fbsims$ sense transistors to make sure that the appropriate phase of channel stresses and strains are generated. The model is incomplete without the inclusion of parasitic capacitances and resistances associated with the metal traces, as well as self and coupling capacitances of each net. The Calibre xACT\textsuperscript{\texttrademark} tool is used to perform the parasitic extraction on the layout of the fRBT device. The extent of layout corresponds to the de-embedding plane used in the RF measurements to extract device characteristics. Once a netlist is generated after parasitic extraction, the standard transistors are replaced with their $fbsim$ counterparts while keeping a track of the location of the each. The netlist is also augmented to include the fRBT Body module and the mechanical connections between each of the components in the model. Owing to the high frequency of operation, EM extraction can also be performed for greater accuracy. Since EM analysis is comparatively computationally expensive and difficult to integrate with the rest of the model, it was excluded in this iteration of the fRBT model development. In the testbench for the complete extracted fRBT model depicted in Fig. \ref{fig10}(b), ideal baluns are used to handle the differential output and input signals for the ease of calculating the differential electromechanical transconductance $g_{m,em}$. The drive and sense transistors are biased using large inductors to mimic the biasing through bias-Ts in the measurement setup. The output  is maintained at 0V DC bias to extract the output current in the simulation. 

\section{Results and Discussion}

\begin{figure*}[ht]
\centering
\includegraphics[scale=0.23]{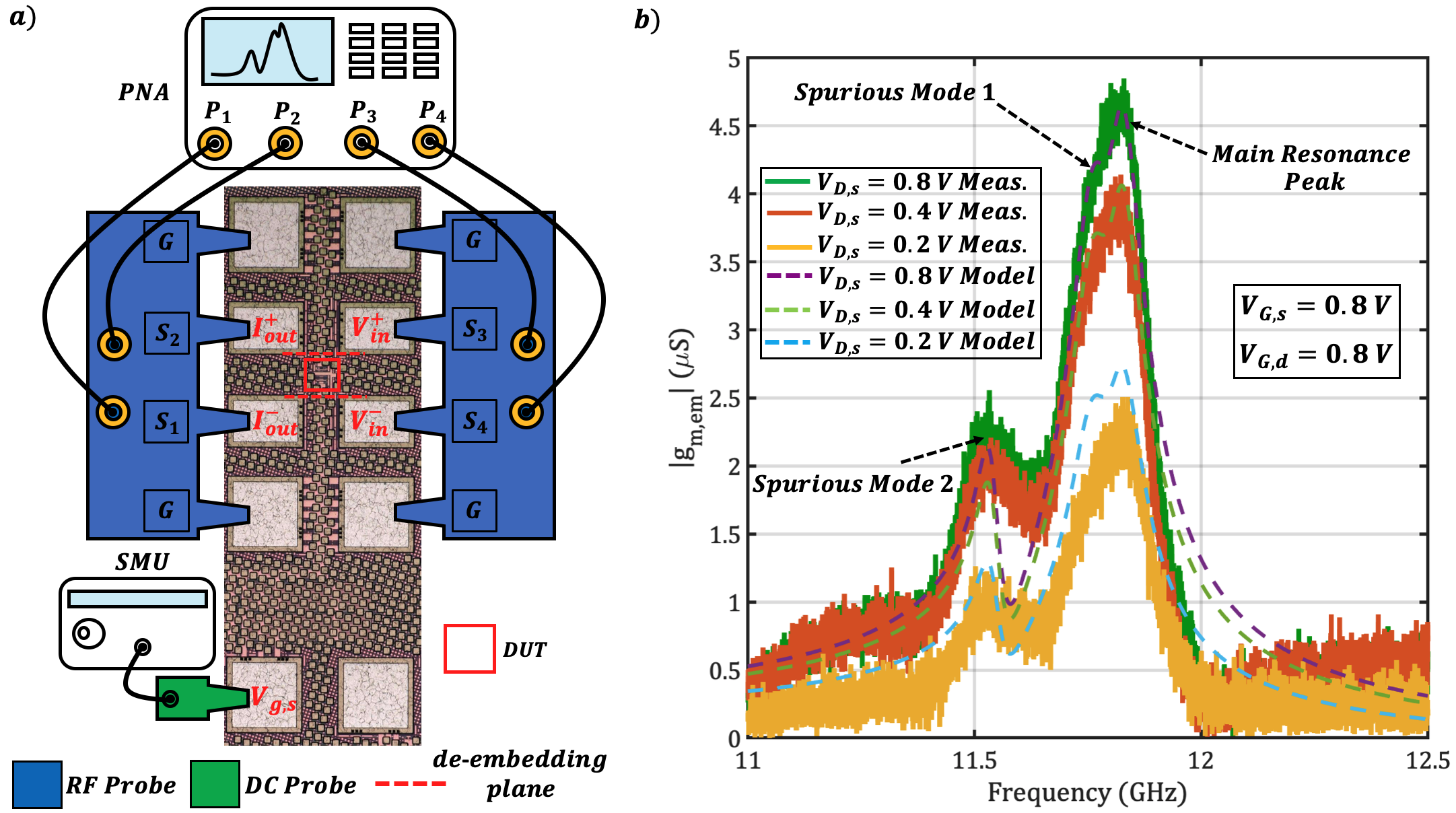}
\vspace{-2mm}
\caption{(a) Measurement setup for the fRBT DUT depicting the GSSG and DC probe landing. The quantities $V_{in}^{+}$, $V_{in}^{-}$, $I_{out}^{+}$ and $I_{out}^{-}$ represent a combination RF signals and DC levels at the different ports of the DUT.  (b) Comparison of measured data from the fRBT device to the extracted fRBT model with variation in the sense transistor drain voltage $V_{D,s}$ for fixed drive and sense gate bias voltage of 0.8 V.}
\label{fig12}
\vspace{-4mm}
\end{figure*}

On-chip RF measurements of the fRBT devices were conducted \cite{anderson2021xband} using the setup shown in Fig. \ref{fig12}(a). Prior to the measurement, Hybrid SOLR-LRRM\cite{SOLR-LRRM} calibration is carried out using impedance standard substrates. Once the quality of the calibration was ascertained to be within recommended tolerances, the DUT was connected to a Agilent\textsuperscript{\textregistered} Parametric Network Analyzer (PNA) through Cascade\textsuperscript{\textregistered} GSSG probes landed on the DUT probe pads. The internal bias-Ts of the PNA were connected to two Source-Measure Units (SMUs), one each for the input gate drive bias (S\textsubscript{3} and S\textsubscript{3}) and for the output sense transistor drain bias (S\textsubscript{1} and S\textsubscript{2}). A third SMU provided the sense transistors' gate bias using a DC probe. An input signal level of -10 dBm was used for all of the measurements. Standard 4-port S-parameter measurement was performed at each bias point and the resulting single-ended parameters were converted to mixed-mode or differential parameters. Conversion of the differential S-parameters to differential Y-parameters provides extraction of the electromechanical transconductance of the fRBT DUT:

\begin{equation} \label{eq:gmem}
    |g_{m,em}| = |Y_{21dd}-Y_{12dd}|
\end{equation}

where $Y_{21dd}$ and $Y_{12dd}$ are differential $Y$-parameters. Open and short structures present on-chip are used for de-embedding parasitic elements associated with pads and routing up to the de-embedding plane specified in Fig. \ref{fig12}(a). This includes bias-dependent capacitance of electrostatic discharge (ESD) diodes necessary to protect the fRBTs. In practice, devices would be routed directly to adjacent circuits in low-level metal layers, irradicating the need for ESD diodes and extensive routing for each device.  

\begin{figure}[ht]
\centering
\includegraphics[scale=0.19]{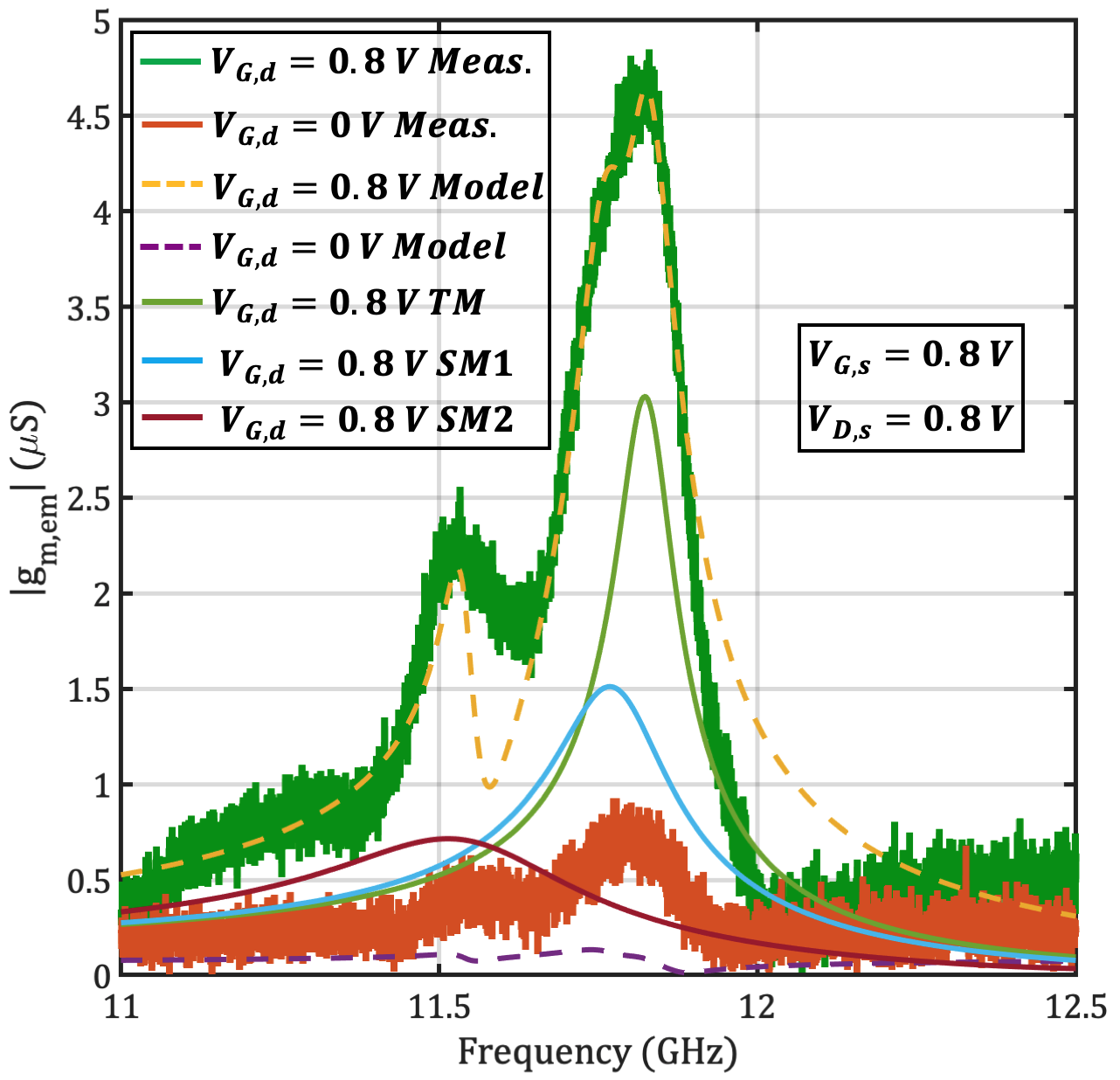}
\vspace{-2mm}
\caption{Variation of the fRBT $g_{m,em}$ response with drive gate bias $V_{G,d}$ for sense transistor drain ($V_{D,s}$) and gate ($V_{G,s}$) biased at 0.8 V. Along with the measured data and the extracted model $g_{m,em}$, the TM, SM1 and SM2 curves represent the individual resonant characteristics of the target mode, spurious mode 1 and 2 respectively. The three modes collectively combine to give the overall response matching the measured data.}
\label{fig13}
\vspace{-4mm}
\end{figure}

\vspace{3mm}
In the first set of measurements, the drive MOSCAPs are biased into inversion with $V_{G,drive} = 0.8 V$. The sense transistor gate bias is fixed at $V_{G,s} = 0.8 V$ and the drain bias voltage is varied to assess the performance in different regions of operation. From Fig. \ref{fig12}(b), we see that as $V_{d,s}$ is lowered from $0.8 V$ (saturation regime) to $0.2 V$ (linear regime), the peak $g_{m,em}$ reduces in accordance with a reduction in DC drain current. In each of the three measured curves, two spurious modes in the vicinity of the desired resonance peak can be observed, one of which is merged in close proximity with the target mode. This likely results from the finite nature of the resonance cavity in the $x_{2}$ direction. Harmonics generated by transverse modes such as these cannot be captured by unit-cell-based 3D FEM models, and would require extensive modeling of the complete 3D structure which is currently computationally prohibitive. The Q of the main resonance peak is extracted and fed back to the fRBT model as an initial point for the final model fit. As can be seen, the model captures the presence of the two spurious modes along with the targeted mode. Some discrepancies are observed between the measured response and the model in the region between the spurious modes and in a transmission zero beyond the measured main resonance. These can be attributed to limitations of the parasitic extraction which excludes some of the coupling capacitance across the fRBT structure. The model shows commensurate variation with $V_{d,s}$ as the measured data, validating that the current sensing mechanism modeling is able to accurately capture the effect of bias variation.  

\vspace{3mm}
An important feature of electrostatically-driven electromecahnical devices is the ability to control the strength of the drive transducer with DC bias. The resonance is not completely attenuated since a capacitance still exists when the bias goes down to $0$ $V$, which results in transduction. As can be seen from the corresponding model curve for the $V_{G,d} = 0$ $V$ bias, the $g_{m,em}$ is attenuated.

\begin{figure}[t]
  \centering
  \begin{tabular}{@{}c@{}}
    \includegraphics[scale=0.14]{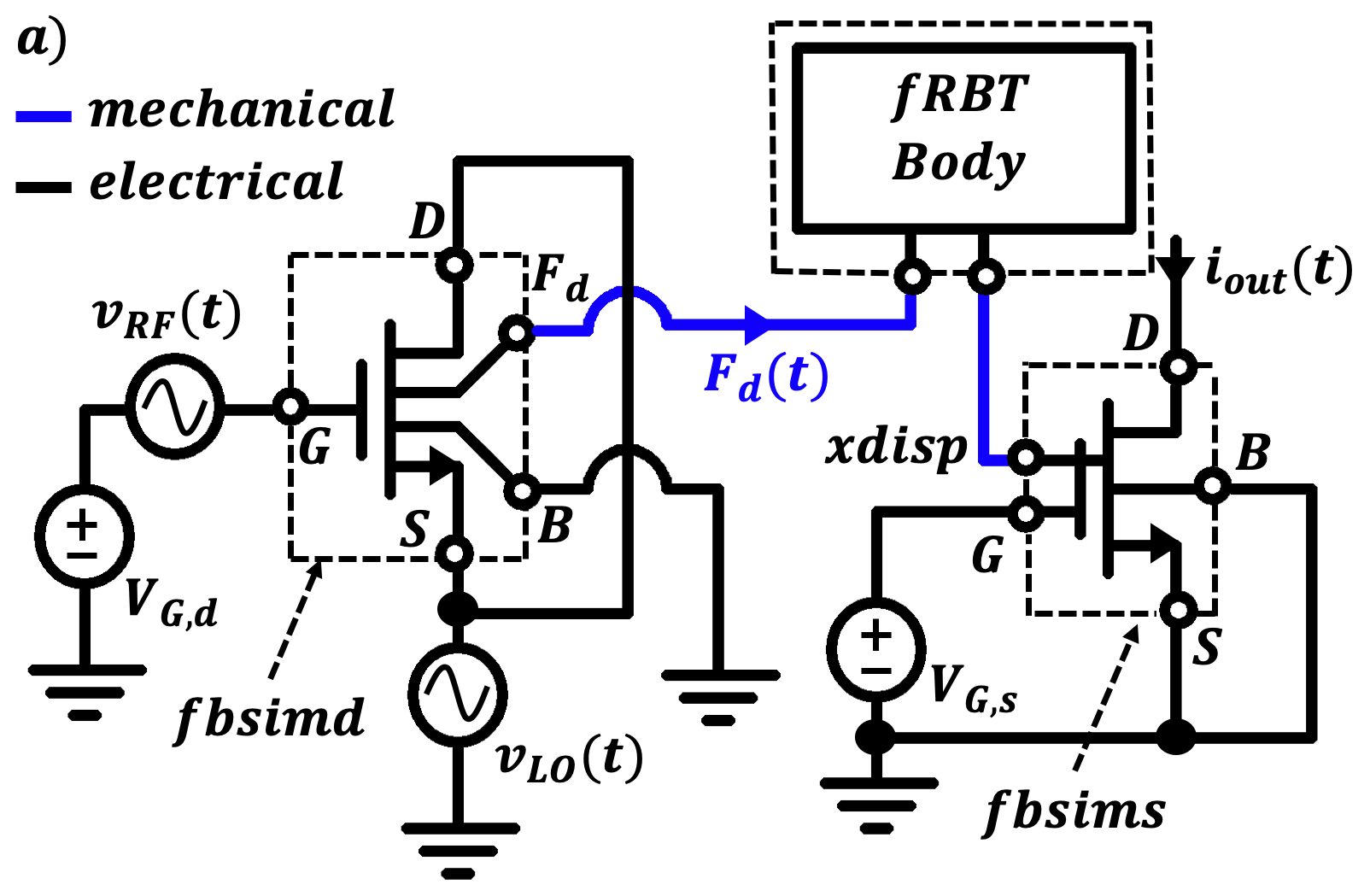} \\
  \end{tabular}


  \begin{tabular}{@{}c@{}} 
    \includegraphics[scale=0.18]{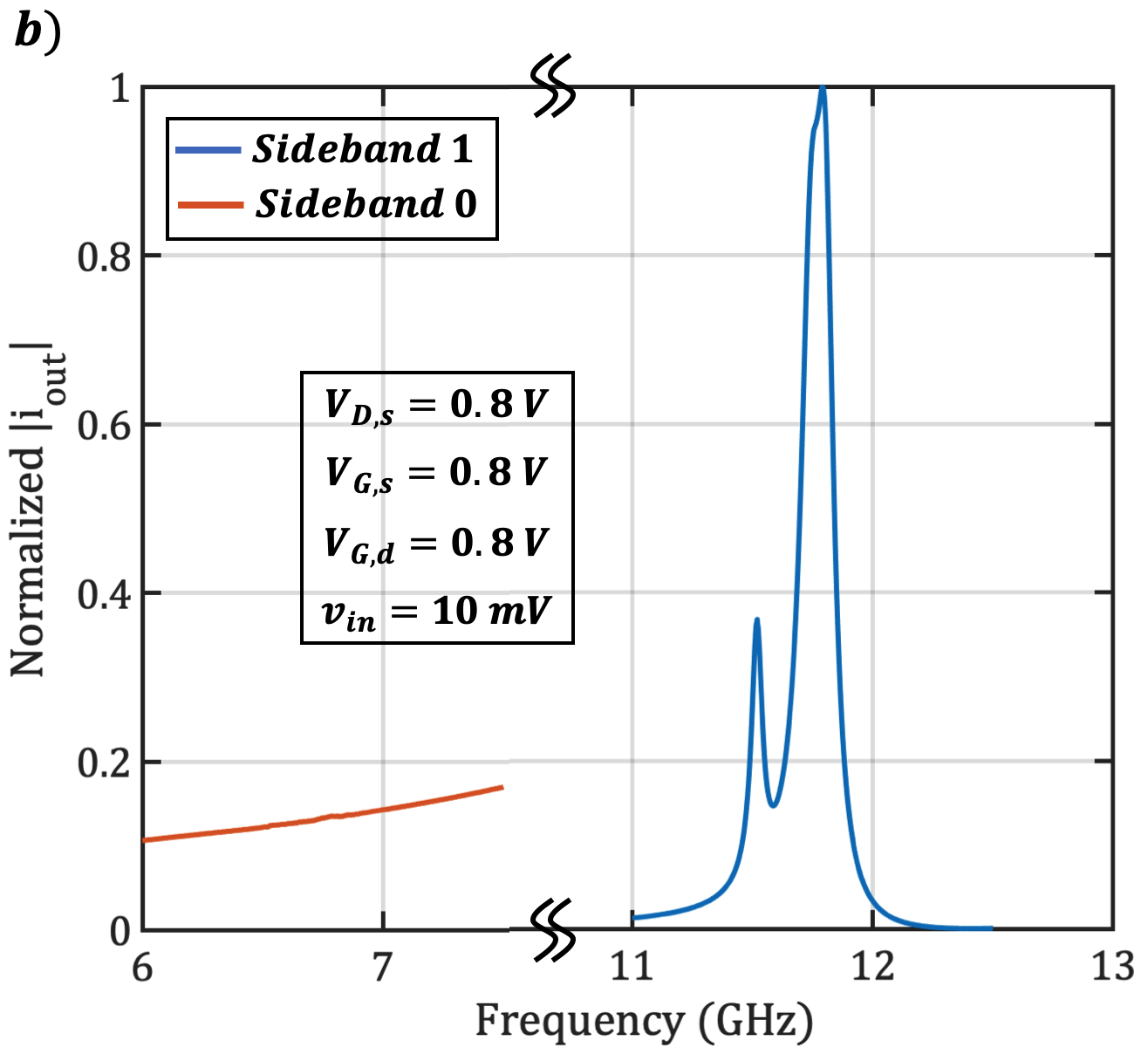} \\
  \end{tabular}
  \caption{(a) PSS and PAC testbench circuit schematic for emulating the mixer measurement setup typically used for resonators embedded in high feedthrough. (b) PAC analysis results showing the presence of the resonance characteristic of the output current in the sideband corresponding to $f_{0}=f_{RF}+f_{LO}$.}\label{fig14}
  \vspace{-2mm}
\end{figure}

The developed fRBT model is also compatible with Periodic Steady State (PSS), Periodic AC (PAC) and Harmonic Balance simulation scenarios. Since the model is designed to capture the nonlinear mechanisms inherent to the device, we can use these simulation techniques to verify functionality. An RF/LO-based mixing measurement technique is used to extract the performance of resonators embedded in high feedthrough \cite{RFLOmeas}. In this method, along with an RF signal $v_{RF}$ that is applied at the device input, a lower frequency signal $v_{LO}$ is superimposed on the bias voltage. The frequencies of the RF and LO signals are set such that $f_{0}=f_{RF}+f_{LO}$ where $f_{0}$ is the resonance frequency of the resonator. Even though frequencies different from the $f_{0}$ are applied to the resonator terminals, due to the nonlinear electromechanical transduction based up-conversion mixing, a mechanical force is generated at the resonance frequency $f_{0}$. To test the functionality of the fRBT model under these nonlinear simulation conditions that are frequently used for oscillator/filter designs, a simulation setup emulating resonator mixing measurements is created as shown in Fig. \ref{fig14}(a). An RF signal $v_{RF}$ of frequency $f_{RF}$ is applied to the gate of the drive transistor along with the gate bias $V_{G,d}$ of 0.8 V. Instead of clamping the source-drain to ground, a LO signal $v_{LO}$ of frequency $f_{LO}$ is applied as shown. PSS and PAC simulation results corresponding to the output current are obtained as shown in Fig. \ref{fig14}(b). It can be seen that, in the sideband corresponding to the frequencies 11-12.5 GHz the resonance characteristic is exhibited while in the 6 to 7.5 GHz frequency range it is not. This example confirms that the device model is able to capture the effect of MOSCAP drive nonlinearity as ascertained by the PSS-PAC simulation results. 

\section{Conclusion}

A compact model for an 11.8 GHz Fin Resonant Body Transistor fabricated in a commercial 14nm Fin-FET process has been presented, which captures for the first time all prominent device characteristics including electrical and mechanical nonlinearities. Analysis has also been provided of the unique nature of the phononic dispersion in the device owing to the presence of the BEOL PnC, via a theoretical framework and FEM simulations. An Eigenfrequency-based methodology amenable to dielectric transduction in the fRBT for the extraction of equivalent mechanical circuit parameters for the main resonant cavity has also been described, and can be readily adapted to alternate device geometries and resonance modes. The fRBT model presented in this work is fully parameterized and flexible allowing for the addition of more drive/sense transistors, and also captures all major sources of device non-linearity. The model is also completely integrated within the simulation framework alongside the foundry-supplied PDK for the technology. Future improvements to the model include temperature variation and more accurate parasitic extraction. This large-signal model enables direct integration with interface circuitry, necessary for design of RF and mm-wave oscillators and provides more accurate predictions of overall performance during the design and simulation phase of systems employing the fRBT for electromechanical signal processing.

\ifCLASSOPTIONcaptionsoff
  \newpage
\fi



\bibliographystyle{IEEEtran}
\bibliography{fRBT_Modeling.bib}
%

%

\begin{IEEEbiography}[{\includegraphics[width=1in,height=1.25in,clip,keepaspectratio]{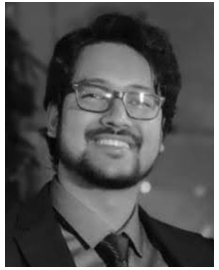}}]{Udit Rawat} received the B.Tech (Hons.) degree from Uttarakhand Technical University, India in 2012 and M.S. by Research degree from Indian Institute of Technology Madras, Chennai, India in 2015. He is currently pursuing a Ph.D. degree with the Department of Electrical Engineering, Purdue University, West Lafayette, IN, USA. From 2015 to 2017 he worked as a Digital Design Engineer at Intel Corporation, Bangalore, India. He is currently a research assistant with the Hybrid MEMS Research group, Purdue University. His research interests include design, fabrication and modeling of RF MEMS resonators as well as their integration in standard CMOS technology. He is also interested in the design of unreleased CMOS integrated gyroscopes and other physical sensors along with their interface circuit design.
\end{IEEEbiography}

\begin{IEEEbiography}[{\includegraphics[width=1in,height=1.25in,clip,keepaspectratio]{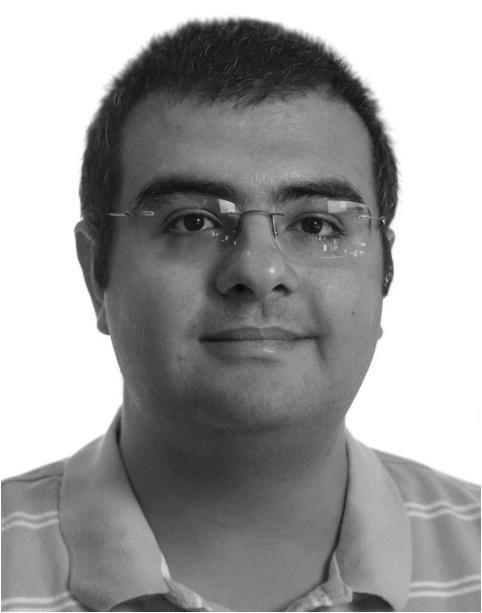}}]{Bichoy Bahr} received the BSc and MSc degrees with honors in 2008 and 2012, both in electrical engineering, from Ain Shams University, Cairo, Egypt. He received the PhD degree from Massachusetts Institute of Technology (MIT) in 2016. He has been with the HybridMEMS group at MIT from 2012 to 2016, where he developed unreleased CMOS-MEMS resonators and monolithic GaN MMIC oscillators. Dr. Bahr joined Texas Instruments' Kilby Labs, Dallas, TX in 2016. He has been elected to TI's prestigious tech Ladder as Member Group, Technical Staff (MGTS) in 2019. His research interests include MEMS resonators, phononic crystals, oscillators,  ultrasonic transducers, and numerical optimization. 
\end{IEEEbiography}

\begin{IEEEbiography}[{\includegraphics[width=1in,height=1.25in,clip,keepaspectratio]{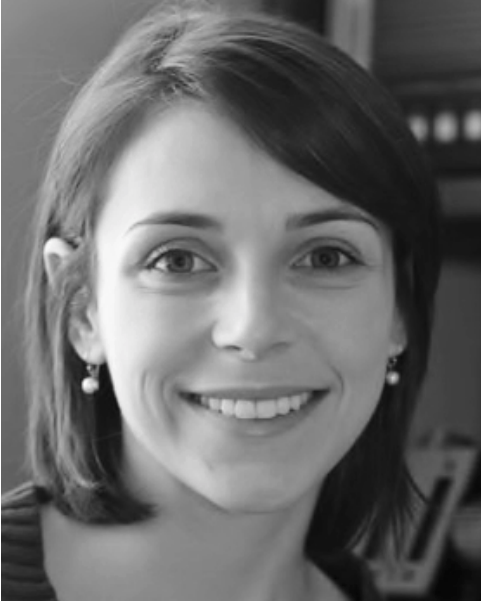}}]{Dana Weinstein} 
received her B.A. in Physics and Astrophysics from University of California - Berkeley in 2004 and her Ph.D. in Applied Physics in 2009 from Cornell, working on multi-GHz MEMS. Prior to joining Purdue as an Associate Professor in Electrical and Computer Engineering in 2015, Dr. Weinstein joined the Department of Electrical Engineering and Computer Science at MIT as an Assistant Professor, and served as an Associate Professor there between 2013 and 2015. Dana is now a Purdue Faculty Scholar and serves as Associate Dean of Graduate Education in the College of Engineering. She is the recipient of the Transducers Early Career Award, NSF CAREER Award, the DARPA Young Faculty Award, the Intel Early Career Award, and the IEEE IEDM Roger A. Haken Best Paper Award.
\end{IEEEbiography}




\end{document}